\documentclass[11pt]{article}
\usepackage{authblk}

\usepackage[numbers,sort&compress]{natbib}
\usepackage{amssymb}
\usepackage{amsmath}
\usepackage{mathtools}
\usepackage{booktabs}
\usepackage{etoolbox}
\usepackage{comment}
\usepackage{microtype}
\usepackage{graphicx}
\usepackage[T1]{fontenc}
\usepackage{lmodern}
\usepackage[margin=1in]{geometry}

\newcommand{\Var}{\mathrm{Var}}

\renewcommand{\d}{\mathrm{d}}

\newcommand{\trp}{{^\top}} %

\newcommand{\subscr}[2]{#1_{\textup{#2}}}
\newcommand{\supscr}[2]{#1^{\textup{#2}}}

\newcommand\oprocendsymbol{\hbox{$\square$}}
\newcommand\oprocend{\relax\ifmmode\else\unskip\hfill\fi\oprocendsymbol}

\newtoggle{bibdoi}
\newtoggle{biburl}
\makeatletter
\newsavebox{\bib@url}
\newsavebox{\bib@doi}

\undef{\APACrefURL}
\undef{\endAPACrefURL}
\undef{\APACrefDOI}
\undef{\endAPACrefDOI}

\title{A Normative Theory of Decision Making From Multiple Stimuli:\\ The Contextual Diffusion Decision Model}
\author[1]{Michael Shvartsman}
\author[2]{Vaibhav Srivastava}
\author[3]{Narayanan Sundaram}
\author[1]{Jonathan D.\ Cohen}
\affil[1]{Princeton Neuroscience Institute}
\affil[2]{Michigan State University}
\affil[3]{Intel Corporation}

\begin{document}
\maketitle

\begin{abstract}
  The dynamics of simple two-alternative forced-choice (2AFC) decisions are well-modeled by a class of random walk models \cite[e.g.][]{Laming1968,Ratcliff1978,Usher2001,Bogacz2006}. However, in real-life, even simple decisions involve dynamically changing influence of additional information. In this work, we describe a computational theory of decision making from multiple sources of information, grounded in Bayesian inference and consistent with a simple neural network. This Contextual Diffusion Decision Model (CDDM) is a formal generalization of the Diffusion Decision Model (DDM), a popular existing model of fixed-context decision making \cite{Ratcliff1978}, and shares with it both a mechanistic and a probabilistic motivation.  Just as the DDM is a model for a variety of  simple two-alternative forced-choice (2AFC) decision making tasks, we demonstrate that the CDDM supports a variety of simple context-dependent tasks of longstanding interest in psychology, including Flanker \cite{Eriksen1974}, AX-CPT \cite{Servan-Schreiber1996}, Stop-Signal \cite{logan1984ability}, Cueing \cite{Posner1980}, and Prospective Memory paradigms \cite{Einstein2005}. Further, we use the CDDM to perform a number of normative rational analyses exploring optimal response and memory allocation policies. Finally, we show how the use of a consistent model across tasks allows us to recover consistent qualitative data patterns in multiple tasks, using the same model parameters.  %
  \end{abstract}

\maketitle

\section{Introduction}

Even the simplest decisions (e.g., \ recognizing a stoplight as red or green while driving) are made based on complex and multidimensional informational input, consisting of both multisensory perception and the content of multiple different memory systems. The ability to make such decisions in a flexible, context-dependent manner is at the core of the construct of \emph{cognitive control}, the ability to adaptively change behavior in response to changing circumstances and task goals.

Approaches to understanding the processes underlying such decision making tend to model a single information source, treating the others as static \cite[e.g.][]{Lositsky2015,Sheppard2013}. In real-world settings, however, there are often multiple dynamically changing sources of information that are relevant to a decision, whether from multiple perceptual streams (e.g.\ noticing a crossing pedestrian while recognizing the stoplight) or from perception and memory (e.g.\ reading signs while recalling directions). Other approaches focus on modeling an underlying mechanistic process \citep[e.g.][]{Botvinick2001,Braver1995}, but are not tractable to closed-form normative analysis. There have been some efforts to provide such formal grounding, but with additions in complexity that themselves prevent analysis of full response time (RT) distributions, providing only data fits and analysis at the level of mean RTs (e.g. \cite{Norris2009,Lewis2013}, though for exceptions see \cite{Liu2009,McMillen2006}).

In our work, we describe the Contextual Diffusion Decision Model (CDDM), which reflects the dynamics of multiple information sources, is grounded in optimal statistical inference, and is simple enough to provide a formal analytic understanding of the relationship between response times and relevant model parameters. We begin with the DDM \citep{Ratcliff1978}, which focuses on two-alternative forced-choice (2AFC) decision tasks, and extend it to encompass the processing of an additional stimulus. We refer to the second stimulus as the \emph{context}, because it determines what task to perform (i.e., the relevant set of stimulus-response mappings), in addition to the imperative stimulus used in standard 2AFC tasks that we refer to as the \emph{target}). We show that the CDDM is a generalization of the standard DDM, a highly influential model of single-stimulus decision making, that models response time as the first passage time of a two-boundary one-dimensional Wiener process with drift. Specifically, the CDDM reduces to the DDM in the limiting cases where one stimulus is known or uninformative (but crucially, not in cases where the contextual stimulus affects the task response).

Using this framework, we perform a number of normative analyses, contrasting optimal and fixed-threshold policies in both Flanker and AX-CPT, demonstrating how previously-proposed explanations for the Flanker effect can emerge as a normative consequence of different task constraints, and providing a new normative explanation of memory decay as a way of mitigating against encoding noise in the AX-CPT. Finally, we validate the generality of the overall framework by showing that it can recover qualitatively correct patterns of behavior in different tasks under a single parameterization derived from prior work. 

While we focus on a relatively simple type of context-dependent processing, we show that the CDDM provides insights into how more ecologically realistic, dynamic multi-context decisions differ qualitatively and formally from single-context decisions. Furthermore, we show that some heuristics, previously used to improve data fits in analyses of standard single-context 2AFC tasks, emerge naturally from our formulation.

The article proceeds as follows: We begin with a brief review of the conventional fixed-context DDM, and how it can be motivated both from a statistical and dynamical systems perspective. Next, we describe our generalization of the DDM to context-dependent decision making that is similarly motivated from both statistical and dynamical systems perspectives, and can be used to model a variety of context-dependent tasks. Then, we describe simulation-based insights into the model's behavior by showing how response time distributions are affected by parameter changes. We also analyze bounded-optimal attention allocation and response policies in this model and show how, for a number of phenomena, previously provided qualitative explanations arise as optimal consequences of different constraints. Finally, we demonstrate the generality of the framework by producing qualitatively reasonable predictions in different cognitive control tasks using a single set of parameters. 

\section{DDM as a unifying perspective on decision making}\label{sec:ddm}

The DDM serves as a unifying perspective on simple, two-alternative decision making in part due to its motivation both as: a normative statistical procedure, in which humans or animals observe a stimulus until the amount of information accrued is optimized for decision making \citep[e.g.][]{Bogacz2006}; and an abstraction of neural population activity in which firing rates rise to a threshold \citep[e.g.][]{shadlen2001neural}. We briefly review both of these perspectives, highlighting aspects that undergird our formulation of the CDDM. 

\subsection{Statistical motivation}

The DDM can be derived as a continuum limit of the sequential probability ratio test (SPRT; \cite{Wald1948}), the optimal sequential statistical test of simple hypotheses. 
The test assumes that the decision maker is faced with a sequence of i.i.d.\ observations $y_1, \ldots, y_m$ drawn from one of two distributions, associated with hypotheses $H_0, H_1$. Then, the following recurrence relation is computed:
\begin{align}
z_t &= \log \frac{p(y_t\mid H_1) }{ p(y_t\mid H_0)}\\
Z_t &= Z_{t-1} + z_t = \sum_{\tau=1}^t \log \frac{p(y_\tau\mid H_1)}{p(y_\tau\mid H_0)}.
\end{align}

\noindent 
The decision policy is:
\begin{align}
\begin{cases}
 \mbox{Sample }y_{t+1}, & \text{ if }   A < Z_t < B, \\
 \mbox{Choose }H_0, & \mbox{ if } A \le Z_t, \\
 \mbox{Choose }H_1, &\mbox{ if } B \ge Z_t.
\end{cases}
\end{align}

That is, cumulative log-likelihood ratios (LLRs) are computed until the sum crosses one of the two thresholds (one for each hypothesis), at which point the decision is made. This is known to be the most efficient information accumulation process for either a specified amount of time and/or a threshold for confidence \cite{Wald1948}.  Furthermore, an optimal threshold can be derived that maximizes returns per unit time (i.e., reward rate; \cite{Bogacz2006}).  The sequence of LLRs forms a random walk to either the $A$ or $B$ bounds. While this test has been used directly as a model of human decision making \cite[e.g.][]{Stone1960,Edwards1965,Ashby1983,Laming1968,Link1975}, additional analyses are possible by recognizing that in the continuum limit, the log-likelihood ratio evolves as the following stochastic differential equation:
\begin{align}
\d x &= a \d t + \sigma \d W_t,\quad x(0)=x_0,
\end{align}
where $a = \mathbb{E}\left[\log \frac{p(y_t\mid H_1)}{p(y_t\mid H_0)}\right], \sigma^2 = \Var\left[\log \frac{p(y_t\mid H_1) }{ p(y_t\mid H_0)}\right] $, and $W_t$ evolves according to a Wiener process. The predicted response time distribution of the continuous model is the solution to a two-boundary first passage time problem of this process, with the boundaries $A$ and $B$ as in the discrete case. A detailed derivation is given in the appendix of \cite{Bogacz2006}, following \cite{Laming1968} and others. \cite{Ratcliff1978} provides an alternate mechanistic motivation for a Wiener process theory of decision making based on a sequence of parallel memory comparisons, arriving at the same diffusion process. This latter formulation also includes trial-to-trial variability in the values of $a$ and $x_0$, which provides a better descriptive fit to data while departing from normative foundations (though see \cite{Drugowitsch2012, callaway2021fixation} for efforts to bridge the two).

\subsection{Biophysical motivation}
\label{sec:ddm-biophys}

The motivation for the DDM as a biophysically plausible model comes from a variety of neural evidence, including rise-to-threshold activity in neural population firing rates preceding a decision in animal models \cite[e.g.][]{Kira2015,Gold2002,Gold2007Neuro,Resulaj2009,Hanks2015}, connectivity patterns in rat basal ganglia \cite{Bogacz2007NeuralComputation}, and evidence for ramping activity during decision making in neuroimaging \cite{VanVugt2012,Turner2015,ODoherty2007,Krueger2017}. Inspired by such evidence, \cite{Bogacz2006} gives a derivation of the DDM by time-scale separation of a simple neural network, while \cite{Wong2006a} and \cite{Eckhoff2011} give a more formal biophysical treatment.

We briefly review the time-scale separation argument of \cite{Bogacz2006}, that we later extend for the CDDM. We begin with the following linearized mutual inhibition model \citep{Usher2001}, consisting of two neuronal populations $y_1, y_2$ selectively tuned to two stimuli:
\begin{align}
\d y_1 &= (r_1 - \lambda y_1 - \rho y_2)\d t + \sigma \d \mathbb{W}_1\\
\d y_2 &= (r_2 - \lambda y_2 - \rho y_1)\d t + \sigma \d \mathbb{W}_2,
\label{eqn:pooled-inhib}
\end{align}
where  $r_{1,2}$ are drift terms, $\lambda$ is a decay term, $\rho$ is a mutual inhibition term, and $\mathbb{W}_{1,2}$ are uncorrelated Wiener processes. 
This model itself can be derived as a simplification of more realistic biophysical models \cite{wong2006recurrent}. 

It can be shown that the eigenvalues of the system are $-\lambda-\rho$ and $\lambda-\rho$. Since both $\lambda$ and $\rho$ are positive by definition, as both self-inhibition and mutual inhibition get larger (in so-called \emph{tight competition}), $-\lambda-\rho$ becomes substantially larger in magnitude than $\lambda-\rho$. Transforming the system into orthonormal basis gives the following decoupled pair of Ornstein-Uhlenbeck (O-U) processes: 
\begin{align}
\d x_1 = (\lambda-\rho)\left(\frac{r_1-r_2}{(\lambda-\rho)\sqrt{2}} - x_1 \right)\d t+\sigma \d \mathbb{W}_1\\
\d x_2 = (\lambda+\rho)\left(\frac{r_1+r_2}{(\lambda+\rho)\sqrt{2}}-x_2 \right)\d t + \sigma \d \mathbb{W}_2
\end{align}

In the tight competition setting, $\lambda+\rho\ge\lambda-\rho$, and therefore $x_2$ approaches its mean far more quickly than $x_1$. If we focus only on the slow timescale dynamics of $x_1$, we see that they are an Ornstein-Uhlenbeck process. With balanced self- and mutual-inhibition it once more reduces to the DDM with drift $a:=r_1-r_2$ and noise $\sigma$.

\subsection{Beyond 2AFC over simple hypotheses}

The statistical and dynamical system groundings of the DDM bridge levels of abstraction, and have situated sequential sampling (whether the DDM itself or closely related models) as the de facto standard for modeling the dynamics of simple, stationary, two-alternative decisions. 

However, most real-world decisions do not fulfill the conditions in which Wald's SPRT or derivations based on it are optimal, and thereby abrogate its statistical motivation: they are composite, multi-hypothesis, non-stationary, and involve unknown evidence distributions. There is no general computationally inexpensive sequential statistical testing framework that is optimal in any of these cases singly, let alone all together, and applying the SPRT-type threshold policy can range from asymptotically optimal (in the multi-hypothesis case: \cite{Baum1994,Dragalin2000, mcmillen2006dynamics}) to arbitrarily poor (with adversarially selected evidence distributions: \cite{Anderson1960}).  

Finally, SPRT-inspired models have well-known empirical deficiencies even when explaining two-alternative forced-choice decision making. The most notable such deficiency is the prediction of identical distributions for correct and incorrect response times in the case of unbiased initial conditions \cite{Laming1968}, something that has long been known not to be a characteristic of empirical data, for which error RTs can be slower or faster depending on experimental conditions.

A number of efforts have been made to address these challenges, including descriptive extensions of the DDM that no longer have a normative motivation \cite[e.g.][]{Ratcliff1978,Ratcliff1999}, as well as normative extensions for tasks under additional complications such as deadlines, or varying stimulus distributions at the trial level \cite{Frazier2008,Drugowitsch2012, VS-SF-JDC-AS:14c}. Here, we present a potential complement to these extensions. Motivated by the study of cognitive control, we are concerned with the case in which the decision maker needs to combine information from multiple sources, potentially separated across time, in order to make a single binary decision. We focus on the Flanker \cite{Eriksen1974} and AX-CPT \cite{Servan-Schreiber1996} tasks, but the formalism can be applied to Stroop, Cueing, and Prospective Memory paradigms as well \cite{Stroop1935,Servan-Schreiber1996,Posner1980,Einstein2005}.

Finally, it is worth noting that our focus on dynamics distinguishes the work we present from previous work that addresses context-dependent changes in preferences \cite[e.g.][]{Srivastava2012} and internal context updating \cite[e.g.][]{OReilly2006}. Furthermore, the admission of evidence from memory distinguishes it from work on multisensory integration \cite[e.g.][]{Sheppard2013}, and the consistency of framing across tasks distinguishes it from earlier task-specific models \cite[e.g.][]{Yu2009}. Recent work has similarly emphasized the dynamic integration of memory and sensory evidence within sequential sampling models \cite[e.g.][]{bornstein2023associative}. Our work complements and extends this line of research by providing a unified normative account of such dynamics, and showing how they arise as optimal decision policies across multiple task domains.

In addition to formal connections to the models discussed above, our model is also complementary to work on the perception-memory tradeoff on longer timescales in the Prospective Memory paradigm \cite{Einstein2005}, and mechanistic work combining multiple sequential samplers in the ACT-R cognitive architecture \cite{VanMaanen2012}. Finally, it contrasts with related efforts that implicitly address the time-varying influence of an additional stimulus \cite[e.g.][]{Smith1988,Busemeyer1993,Usher2001}, in being explicitly grounded out in formal probabilistic inference.

\section{The theoretical framework: Contextual Diffusion Decision Model (CDDM) as a unifying perspective}

Here, we develop an extension of the DDM that parallels the exposition of the one-dimensional case provided above: We begin with a statistical motivation, reduce it to a 2-D diffusion system, namely the CDDM, and then re-derive the CDDM from a simple neural network. 

\begin{figure}[ht!]
    \centering
    \includegraphics[width=0.5\linewidth]{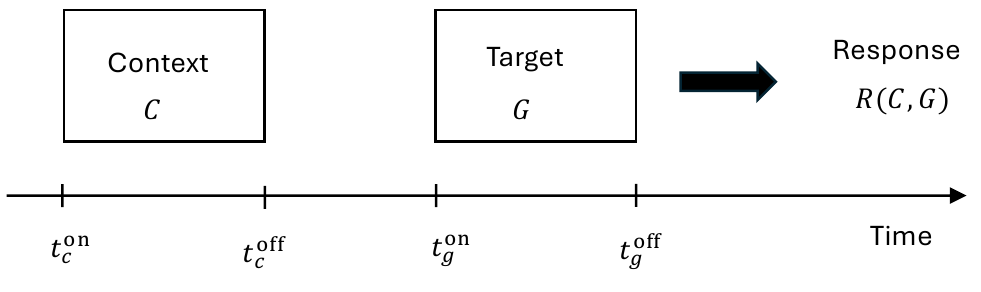}
    \caption{Contextual decision making: A binary context stimulus appears at time $\supscr{t}{on}_c$ and disappears at time $\supscr{t}{off}_c$. A binary target stimulus appears at time $\supscr{t}{on}_c \ge \supscr{t}{on}_c$ and disappears at time $\supscr{t}{off}_g \ge \supscr{t}{off}_c$. The decision maker selects a binary response based on the inferred context and target stimuli. Note that while the figure illustrates the stimuli being fully temporally disjoint, our proposal also applies to the case where they overlap partially or fully in time.}
    \label{fig:context-target}
\end{figure}

We develop the model for the class of tasks shown in Figure~\ref{fig:context-target}, which feature a single contextual stimulus and a single target stimulus that can be concurrent or separated in time, and a single binary decision that must be made. If the contextual stimulus appears before the target, the decision maker can use a simple memory system to retain information about the context.

We assume that the dynamics of multi-stimulus decision making, like single-stimulus decision making, can be understood as a sequential probabilistic inference process, where the decision maker uses sequentially drawn samples to compute the joint posterior probability over stimulus identity over time. Given this posterior probability, we can define a decision rule that determines whether additional samples are taken or the stimulus identity is declared, and the decision is made. This decision rule, together with stimulus onset and offset times, can be used to define different tasks in a consistent mathematical framework. 

\subsection{CDDM: Task specification}
As outlined above, we consider decision making scenarios involving the integration of two stimuli, that we refer to as the \emph{context} and the \emph{target}. 
On a given trial, the context and target each take on one of two binary values, designated  $\{c_0, c_1\}$ and $\{g_0, g_1\}$, respectively.
We denote by $C, G$ random variables representing the possible draws of context and target. 
The only distinguishing characteristic of the target is that, under some conditions, it can appear and disappear before the target.

We denote by $\supscr{t_c}{on}$ the time at which the context appears and $\supscr{t_c}{off} > \supscr{t_c}{on}$ the time at which it disappears, and likewise $\supscr{t_g}{off} > \supscr{t_g}{on}$ the times at which the target appears and disappears. We restrict that $\supscr{t_c}{on}\le \supscr{t_g}{on}$ (i.e.\ the context is always the stimulus that appears first). The onsets and offsets define different tasks, for example, if $\supscr{t_c}{on}=\supscr{t_g}{on}$ and $\supscr{t_c}{off}=\supscr{t_g}{off}$, the task is purely perceptual, whereas otherwise the context must be retained in memory. 

In addition to stimulus onset and offset times, a task is also defined by a reward structure that maps from stimulus-response pairs to goals or rewards. Since we focus on decision making rather than learning, we assume that the reward structure is known to the decision maker. Furthermore, the reward structure can be described by a response rule $R(c_*,g_*)\rightarrow\{r_0,r_1\}$ that deterministically maps from stimulus identity to a specified response. The agent has a policy $\pi(P(C, G))$ that maps from its belief of the true context and target $P(C, G)$ to actions that include sampling actions (that acquire additional information) and response actions (that end the trial). We assume the policy is deterministic. 

This framework captures several prototypical tasks, as discussed below. 

\paragraph{Flanker task} In this task~\cite{Eriksen1974}, a binary target stimulus must be identified in the presence of a context stimulus that can be the same as (congruent with) or different from (incongruent with) the target. Both of the stimuli appear at the same time, and the response rule is independent of the context. This is structurally homologous to other interference paradigms, such as the Stroop task \cite{Stroop1935}. 

\paragraph{AX-CPT task} In this task, a binary context stimulus precedes a binary target stimulus. The binary response rule takes one value if the context and target match and takes the alternative value, otherwise. 
This also corresponds to other tasks, such as match-to-sample \cite{Simola2010}. 

\paragraph{Cueing task} In this task~\cite{Posner1980}, a binary target stimulus must be identified just as in the Flanker case, but in this case the context stimulus (which again can be congruent or incongruent) appears \emph{before} the target rather than concurrently with it. The response rule remains independent of the context. 

\paragraph{Prospective Memory task} In the perspective memory task \cite{Einstein2005}, both the context and the target stimuli appear  concurrently. The binary response rule takes one value for one of the contexts, irrespective of the target, while for the other contexts, the response rule depends on the target. A similar structure exists in the asymmetric AX-CPT \cite[e.g.][]{Henderson2012}. 

We selected these four task paradigms because collectively, they have been used in a remarkably broad range of studies of basic cognitive function examining perception, attention, cognitive control, timing, and response selection. Here, we seek to unify these by noting that they all reflect the use of context in otherwise simple decision making tasks (here, simplified to 2AFC versions of those tasks). More specifically, these four paradigms can be seen as varying along two critical dimensions:  1) the timing of the context, either concurrent with the target (Flanker and AX-CPT) or antecedent to it (Cueing and Prospective Memory); and 2) the context-dependence of the stimulus-response mapping, either dependent (AX-CPT and Prospective Memory) or independent (Flanker and Cueing). This provides a 2x2 task design space in which the interaction between dynamics of processing and context-dependence can systematically be examined.

\subsection{CDDM: Inference model}

We define an abstract \emph{context sensor} and \emph{target sensor} selectively tuned to context or target information, respectively. We first focus on conditions in which the context and target appear concurrently (Flanker and Prospective Memory tasks) and then discuss the more general case. 

\medskip 

\noindent 
\textbf{Concurrent Context and Target Stimuli.}
Let $y_k^C$ and $y_k^G$ be the samples drawn from the context and target at discrete time $k$. 
We assume that the decision maker integrates the noisy samples it receives in a Bayesian fashion.  With i.i.d.\ observations of context and target, this posterior probability distribution contains all the information the decision maker needs to make a decision. Let:
\begin{itemize}
\item $l_k(c_i,g_j)=P(y_k^C,y_k^G\mid C=c_i, G=g_j)$ denote the agent's likelihood function; %
\item $p_{k}(c_i,g_j)=P(C=c_i, G=g_j)$ denote the agent's belief at time $k$ that the true context and target are $c_i$ and $c_j$, respectively.  %
\end{itemize}
At time $k$, the decision maker receives samples from both the context and the target and uses Bayes' rule to compute the posterior probability %
\begin{align}
p_{k+1}(c_i,g_j) &= \eta_k l_k(c_i,g_j)p_{k}(c_i,g_j), \quad i, j \in \{0,1\}, 
\label{eqn:external-bayesrule}
\end{align}
where $\eta_k$ denotes the normalization constant that ensures that the posterior probabilities sum to 1. With two possible identities for each of $C,G$, the agent's belief resides in the 3-simplex (i.e.\ it needs to represent the probability of three stimulus pairs, with the fourth fully determined from the remaining three). This makes the above update a three-variable discrete-time difference equation. We model the likelihoods as uncorrelated, i.e.\ $l_k(c_i,g_j)=l_k(c_i)l_k(g_j)$, where $l_k(c_i)$ and $l_k(g_j)$ are marginal likelihoods of the context and target, respectively. For example, this may be the case when the decision maker adopts a na{\" i}ve Bayes approach by integrating context and target information while ignoring their potential correlation. 

In accord with the prototypical tasks discussed earlier, we assume that every context-target pair maps deterministically to a response. In other words, given the set $\Gamma =\{\{c_0, g_0\}, \{c_0, g_1\},\{c_1, g_0\}, \{c_1, g_1\} \}$ of all context-target pairs, there exist subsets $\Gamma_0, \Gamma_1 \subset \Gamma$ such that every context-target pair in $\Gamma_i$ maps to response $r_i, \ i \in \{0,1\}$. In this case, we can calculate the log-likelihood ratio of the responses $r_1$ and $r_0$ being correct by
\begin{equation}\label{eq:response-LLR}
   Z_k= \log \frac{P(R=r_1| y^C_{1:k},y^G_{1:k} )}{P(R=r_0| y^C_{1:k},y^G_{1:k} )} = \log \frac{\sum_{\{c_i, c_j\}\in \Gamma_1} \prod_{\tau=1}^k l_\tau(c_i) l_\tau(g_j)}{\sum_{\{c_i, c_j\}\in \Gamma_0} \prod_{\tau=1}^k l_\tau(c_i) l_\tau(g_j)}.  
\end{equation}

Note that the log-likelihood ratio $Z_k$ is a sufficient statistic to compute the posterior probability of a response being correct. By dividing the numerator and denominator of the right-hand side of equation~\eqref{eq:response-LLR} by $\prod_{\tau=1}^k l_\tau(c_0)l_\tau(g_0)$, we can write $Z_k$ as a function of $\prod_{\tau=1}^k \frac{l_\tau(c_1)}{l_\tau(c_0)}$, $\prod_{\tau=1}^k \frac{l_\tau(g_1)}{l_\tau(g_0)}$, and prior probabilities. Equivalently, we can write the right-hand side as a function of $\sum_{\tau=1}^k \log \frac{l_\tau(c_1)}{l_\tau(c_0)}$, $\sum_{\tau=1}^k \log \frac{l_\tau(g_1)}{l_\tau(g_0)}$, and prior probabilities. 

To transform this model into continuous time, we take a continuum limit. Informally, this means we subdivide each posterior update from discrete-time $k-1$ to time $k$ into infinitely many increments, which means that the increments converge in distribution to a Gaussian under the central limit theorem. Furthermore, the pair of summed log-likelihood ratio processes converges to a bivariate Wiener process by the multivariate Donsker invariance theorem~\cite{billingsley1995probability}. 
Specifically, in the continuum limit, the two-dimensional random walk induced by $\begin{bmatrix}\sum_{\tau=1}^k \log \frac{l_\tau(c_1)}{l_\tau(c_0)} & 
\sum_{\tau=1}^k \log \frac{l_\tau(g_1)}{l_\tau(g_0)}
\end{bmatrix}^\top $ 
reduces to $\vec{z}(t) = [z_c(t)\ z_g(t)]^\top$ governed by the following stochastic differential equation:
\begin{align}
\d\vec{z} = \vec{a}\d t + \mathbf{S} \d \mathbb{W}_t, \label{eqn:2d-diffusion}
\end{align}
where 
\begin{align*}
\vec{a} &= \begin{bmatrix}a_c \\ a_g\end{bmatrix},\quad \mathbf{SS}\trp := \begin{bmatrix}\sigma_c^2 & \rho \sigma_c\sigma_g \\\rho \sigma_c\sigma_g  & \sigma_g^2\end{bmatrix}, a_s = \mathbb{E}\left[\log \frac{l_\tau(s_1)}{l_\tau(s_0)}\right], \sigma_s^2 = \Var\left[\log \frac{l_\tau(s_1)}{l_\tau(s_0)}\right], s\in \{c, g\},\\ \rho &=\mathrm{Corr} \left[\log \frac{l_\tau(c_1)}{l_\tau(c_0)}, \log \frac{l_\tau(g_1)}{l_\tau(g_0)}\right], 
\end{align*}
$\mathbb{W}_t$ is the standard two-dimensional Wiener process and $t$ is the continuous time. Importantly, even though this reduction assumes that the agent treats the likelihoods of the context and target as uncorrelated, correlations between the samples of $y^C_\tau$ and $y^G_\tau$ can still yield correlations in $\mathbf{SS}\trp$. 

Given the continuum limits $z_c(t),z_g(t)$, we can also define the continuum limit of the ratio of response likelihoods $Z_k$ defined in equation~\eqref{eq:response-LLR} by
\begin{equation}\label{eq:response-LLR-continuum}
   Z(t)= \log \frac{P(R=r_1|z_c(t),z_g(t) )}{P(R=r_0| z_c(t),z_g(t))},
\end{equation}
where the log-likelihood ratios in~\eqref{eq:response-LLR} have been replaced by their continuum limit. 

\medskip 

\begin{figure}[ht!]
    \centering
    \includegraphics[width=0.8\linewidth]{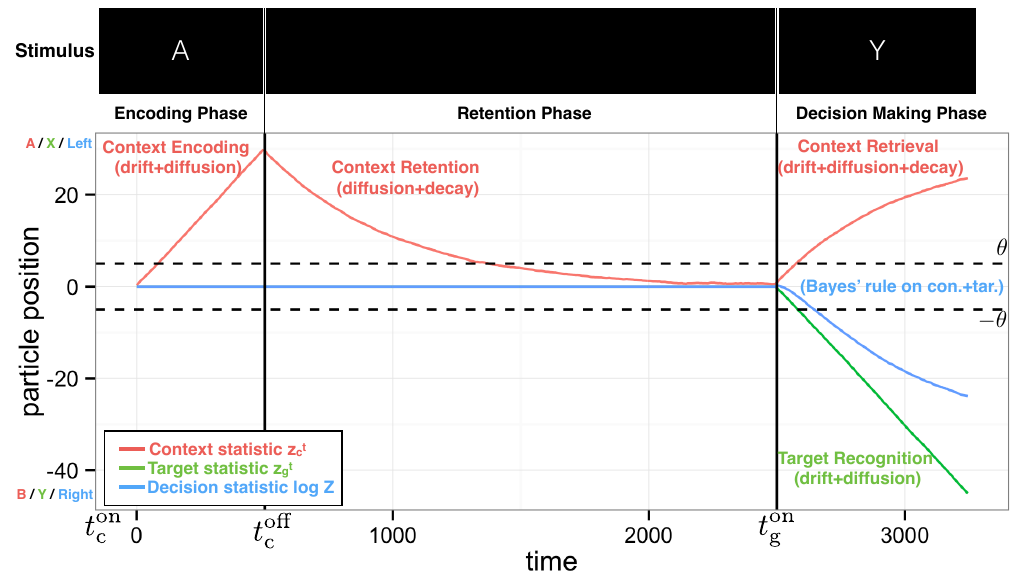}
    \caption{Temporally-separated context and target stimuli illustrated using the example of the AX-CPT task. First, a context stimulus is presented, and the decision maker infers the context. Subsequently, no stimuli are present for a period, and the memory of the context decays. Finally, the target stimulus appears, and the decision maker recalls the inferred context from the memory that is used, together with information from the target, to make a decision. 
    }
    \label{fig:temporally-seperated-stimuli}
\end{figure}

\noindent 
\textbf{Temporally-separated Context and Target Stimuli.}
We next focus on the conditions in which the context and target stimuli are separated in time (e.g., the AX-CPT and Cueing tasks) as shown in Fig.~\ref{fig:temporally-seperated-stimuli}. Here, the context stimulus is presented first that the decision maker must encode. We refer to this as the \emph{encoding phase}. After a specified duration, the context stimulus is removed, and there is a period during which no stimuli are present and the memory of the context decays. We refer to this as the \emph{retention phase}. Finally, the target stimulus appears and the decision maker recalls the inferred context from the memory that is used, together with accumulating information about the target, to make a decision. We refer to this as the \emph{decision making phase}.  

We assume that the participant's prior belief about a stimulus (whether context or target) is stationary until the stimulus appears and then evolves according to our inference process. Thus, in the first phase, when only the context stimulus is present, the inference process can be modeled using standard single-stimuli models such as DDM starting at $\supscr{t_c}{on}$ and initialized according to this prior.

Once the context stimulus disappears from perception at $\supscr{t_c}{off}$, its internal representation is no longer driven by external sensory input. We model the retention of this representation as a zero-mean Ornstein-Uhlenbeck (O–U) process, with decay rate $\lambda$ and noise intensity $c_r^2$, reflecting passive memory decay and internal noise during the retention interval. At the time the target appears, $\supscr{t_g}{on}$, the context must be actively retrieved to inform the decision. We model this retrieval process by introducing a nonzero drift term, with retrieval drift parameter $a_r$, while maintaining the same decay rate. This produces a noisy estimate of the context that evolves dynamically over time. Simultaneously, sensory evidence about the target stimulus begins to accumulate. This target-driven evidence accumulation follows a standard sequential sampling process, with drift determined by the target stimulus. 

Using simple sequential sampling models for memory systems is a common choice on descriptive grounds \citep[e.g.][]{Ratcliff1978,Callaway2023}, and can be thought of as a continuum limit of auto-regressive Temporal Context Models \citep{Howard2002}, or even more directly random context models \citep{Murdock2001,Estes1955,Mensink1988}. An alternate motivation for our O-U process model of memory is the assumption that the memory system is implemented using a leaky competing accumulator model with large leak and mutual inhibition \cite{Bogacz2006}. We provide an alternate new mechanistic motivation for such a model in~\ref{sec:ehrenfest}.

Consistent with the memory model, we add an exponential memory decay term $-\Lambda \vec{z} \d t$ to expression~\eqref{eqn:2d-diffusion}, with $\Lambda$ holding the decay coefficients to get 
\begin{align}
\d\vec{z} = (\vec{a}-\Lambda \vec{z} )\d t+ \mathbf{S} \d \mathbb{W}_t. \label{eqn:2d-diffusion-decay}
\end{align}
More broadly, we will treat $\Lambda$ as a matrix with non-zero off-diagonal terms to capture the potential influence of context and target on one another's sampling. 
We refer to equation~\eqref{eqn:2d-diffusion-decay} as the \emph{contextual diffusion decision model (CDDM)}. 

In this continuous-time formulation, we can additionally reframe our memory system as leading to initial condition variability for the diffusion process. Since we also model the encoding of context as a diffusion process, we can write the position of the context statistic at the end of encoding $z_c^e$ explicitly: 
\begin{align}
z_c^e \sim \mathcal{N}(a_e\Delta t_e, \sqrt{\Delta t_e}),
\end{align}
where $\Delta t_e$ is the duration of encoding. Since retrieval is noisy exponential decay, the position of the statistic at the end of retention can also be written explicitly: 
\begin{align}
\mathcal{N}\left(z_c^e e^{-\lambda\Delta t_r}, \sqrt{\frac{c_r^2}{2\lambda}(1-e^{- 2\lambda\Delta t_r})}\right).
\end{align}

For the case in which context is retained but not retrieved from memory once the target appears, the resultant model is a DDM with initial condition variability \cite{Ratcliff1998}, previously used to fit fast errors in decision making. 
Our analysis provides a possible psychological interpretation for such fast errors: It suggests that data that is better fit using sampling or diffusion models with a variable initial condition may reflect participants' memory of relevant contextual information. Furthermore, it provides a theoretical argument for using a Gaussian rather than the typically used uniform distribution for initial condition variability \cite[e.g.][]{Ratcliff2008}, especially if participants may be using contextual information. Such a model may predict instantaneous decisions (fast guesses) for some parameter settings that reflect strong memory encodings, which may be smoothed out to a multi-modal response distribution if nondecision time variability is included in the model. Importantly, this memory model permits analysis of both external (simultaneous) context (such as in the Flanker and Prospective Memory tasks)
and internal (memory-based) context (as in the AX-CPT and Cueing tasks) in a consistent framing, with the distribution of the initial condition capturing the effect of memory.

\subsection{CDDM: Decision policies}\label{sec:decision policies}

A policy is a mapping from state (in our case the decision maker's beliefs) to actions (making a decision or choosing to sample further). We focus on two classes of deterministic policies: fixed-threshold (FTR) and Bayes risk optimal (BRO) policies. 

\medskip 

\noindent
\textbf{Fixed-Threshold (FTR) Policy:} In fixed-threshold policies sampling from all stimuli is obligatory, and the agent only needs to decide when to stop. Similar to the DDM, the decision maker compares the log-likelihood of a response being correct $Z(t)$ defined in equation~\eqref{eq:response-LLR-continuum} with two thresholds, and selects a response if one of the thresholds is crossed. Specifically, the decision policy is
\begin{align}\label{eq:ftr-rule}
\begin{cases}
 \mbox{Sample both context and target}, & \text{ if }   A < Z(t) < B, \\
 \mbox{Choose response } r_1, & \mbox{ if } Z(t) \ge A, \\
 \mbox{Choose response }r_0, &\mbox{ if } Z(t) \le B,
\end{cases}
\end{align}
where $B <0 <A$ are decision thresholds. In general, such policies are not optimal (though they are optimal for the DDM). Note that the $Z_t = A$ and $Z_t = B$ implicitly define response boundaries in the two-dimensional $\vec{z}$ space.

This heuristic FTR policy is similar to a fixed threshold over their joint density (i.e.\ a 4AFC task where each stimulus pair maps to one response). Such a 4AFC model would be equivalent to the multihypothesis sequential probability ratio test \cite[MSPRT;][]{Tartakovskii1988, Dragalin1999, mcmillen2006dynamics}, an asymptotically optimal test in the sense that it approximates the Bayes-optimal test as accuracy approaches 100\%. However, such a rule may lose even asymptotic optimality guarantees if multiple pairs of context and stimuli map to the same decision, as is the case in many of our motivating tasks, including the Stroop, Flanker, AX-CPT, and Prospective Memory.

We show for specific tasks that $Z(t)$ can be interpreted as a diffusion model with a non-stationary drift, which provides an alternate explanation for the
lack of even asymptotic optimality of FTR policies.  
Nonetheless, this is a sensible heuristic because the space of fixed threshold policies over the response posterior is simple enough to be easy to implement by agents with limited computational resources \citep{Gigerenzer1996}, and low-dimensional enough to explore and optimize quickly by such agents \cite[e.g.][]{Simen2006,Lieder2020}. 
We additionally show that, consistent with it being a continuous-time variant of the model in \cite{Yu2009} when specialized to the Flanker task, it generates plausible behavior patterns. 

In contrast, optimal policies must be defined over the full belief simplex \cite{Dragalin1999}. Presently, a closed-form optimal test is not available, and may not be possible in general. However, the question within cognitive science and neuroscience -- of whether natural agents approximate optimal decision making policies in multi-stimulus settings -- is far from settled, with some recent work suggesting that participants do use or approximate nonstationary policies, specifically ones that either accelerate the sampling rate or lower the threshold as time passes \cite{Drugowitsch2012,Thura2012,Frazier2008}. 
Such policies are optimal if the sampling distribution of the evidence (or equivalently, the observation model) is not known, or in finite-horizon settings. 
Thus, the question of whether fixed threshold policies or richer policies are used in decision making remains an area of active research, and our work engages with both perspectives. 

\medskip 

\noindent
\textbf{Bayes Risk Optimal (BRO) Policy:} 
We illustrate next how optimal policies can be derived from our model. We focus on Bayes-risk optimal (BRO) policies, which are policies that minimize a weighted sum of decision time and error rate, rather than reward-rate optimal (RRO) policies, because we can compute the former by dynamic programming, whereas it is not possible for the latter outside some narrow settings \cite{Dayanik2013}. 

To find the optimal policy for our model by dynamic programming, we re-specify it as a discrete-time partially observable Markov decision process (POMDP; see \cite{Dayan2008} for more on the connections between sequential testing, diffusion, and Markov decision processes). A POMDP is defined by a set of states, a set of actions, a state transition matrix, an observation matrix, and a reward or cost for each state (or state-action pair). The states are the four possible identities of the stimulus pair ($\{c_i,g_j\}, i, j\in\{0,1\}$). 
The actions are the sampling of both stimuli $\subscr{s}{both}$, and the responses $r_0$ and $r_1$. We also add the actions to sample $s_c$ and $s_g$ independently, where this is possible for the decision maker (e.g.\ by shifting attention).
Let the cost of sampling the context be $L_c$, the cost of sampling the target be $L_g$, the cost of sampling both be $\subscr{L}{both}$ (which can potentially be greater or lesser than $L_c+L_g$), and the penalty for making an incorrect response be $\subscr{L}{incrt}$. %

For action $\subscr{s}{both}$, the joint context-target belief-state $\vec{z}(t)$ evolves per the CDDM~\eqref{eqn:2d-diffusion-decay}. For action $s_c$, the context belief-state $z_c(t)$ evolves per the DDM, and the target belief-state $z_g(t)$ remains frozen.  A similar update applies under action $s_g$. For actions $r_0$ and $r_1$, no update is made to $\vec{z}$ and the decision process stops. 

For a given state $\vec{z}(t)$ at time $t$, for action $\subscr{s}{both}, s_c$ and $s_g$, the decision maker incurs cost $\subscr{L}{both}, L_c$, and $L_g$, respectively. 
The decision maker incurs a cost $\subscr{L}{incrt} P_{r_1}(\vec{z})$ for action $r_0$ and a cost $\subscr{L}{incrt} (1-P_{r_1}(\vec{z}))$ for action $r_1$, where given the CDDM state $\vec{z}$, $P_{r_1}(\vec{z}) = \frac{\exp(Z_t)}{1+ \exp(Z_t)}$ is the probability of response $r_1$ being correct.  The cost incurred at any time after taking action $r_0$ or $r_1$ is zero. We solve this POMDP using backward induction with a discretized belief space and a sufficiently large time horizon, though other methods are possible.

The BRO policies, in contrast to FTR policies, allow for selectively tuning decision maker's attention to the context, the target, or both. In this sense, these policies are attention-optimal. We show the differences between fixed thresholds and optimal policies further below, where we discuss specific tasks.

\subsection{CDDM via neural network reduction}

As discussed in Section~\ref{sec:ddm}, the appeal of DDM is that it bridges between statistical perspectives on optimal decision making and their implementation (or at least approximation) in the computations of neural networks \cite[e.g.][]{Usher2001, Bogacz2006}.
Similarly, we show how a neural network can approximate the CDDM by restricting to slower dynamics. 

\begin{figure*}[ht!]
  \includegraphics[width=\textwidth]{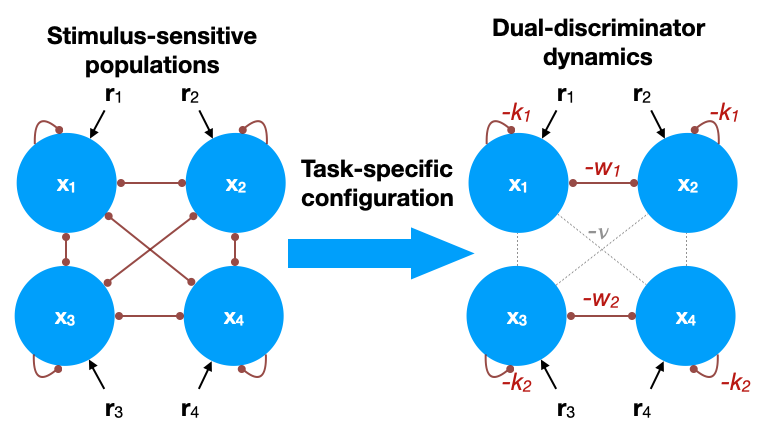}
  \caption{\textbf{Neural network reduction.} \emph{Left:} the full network, consisting of four units each taking one 
  input and all-to-all inhibition. \emph{Right}: if inhibition is much higher between than across pairs, the system can be
  asymptotically approximated as a two-unit system.}
  \label{fig:neural-circuit}
\end{figure*}

We begin with a four-variable linear model with self-inhibition and all-to-all inhibition. Such a model can be obtained by starting with a pooled-inhibition model with an inhibitory pool, and then assuming that the decay of the inhibitory pool is fast relative to the stimulus pools; see for example~\cite{Bogacz2004}. Thus, we can approximate the inhibitory population by its steady-state rate, which is then substituted back into the expression. Each variable in this model reflects the firing rate of a single neural population tuned to one of four context-target stimuli pairs, and inhibits both itself (leak) and all the other populations (competition; see Fig.~\ref{fig:neural-circuit}, left). 

We can configure this set of four populations into two discriminators by reducing the inhibition terms between populations sensitive to stimuli we are not trying to discriminate among (e.g. ,context-target pairs in $\Gamma_0$ or $\Gamma_1$), and increasing the inhibition terms between populations sensitive to stimuli we are trying to distinguish (Fig.~\ref{fig:neural-circuit}, right). We simplify by restricting the mutual inhibition terms to within-discriminator inhibition: $w_1, w_2$ and cross-discriminator inhibition $\nu$. This gives the following model with a block-structured, symmetric dynamics matrix: 

\begin{align}\label{eq:4d-lca}
 \d \vec{y}&= (\vec{r} - \mathbf{A} \vec{y}) dt + %
 \mathbf{C} \d \mathcal{W}_t,
\end{align}
where 
\begin{align*}
\vec{r} = \begin{bmatrix}r_1 \\ r_2 \\ r_3 \\ r_4\end{bmatrix}, \; 
\mathbf{A} = \begin{bmatrix}- k_{1} & - w_{1} & - \nu & - \nu\\- w_{1} & - k_{1} & - \nu & - \nu\\- \nu & - \nu & - k_{2} & - w_{2}\\- \nu & - \nu & - w_{2} & - k_{2}\end{bmatrix}, \; 
\text{ and } 
\mathbf{C}\mathbf{C}^{\top} = \left[\begin{matrix}\sigma_1^2 & \sigma_{12} & \sigma_{13} & \sigma_{14}\\\sigma_{12} & \sigma_2^2 & \sigma_{23} & \sigma_{24}\\\sigma_{13} & \sigma_{23} & \sigma_3^2 & \sigma_{34}\\\sigma_{14} & \sigma_{24} & \sigma_{34} & \sigma_4^2\end{matrix}\right].
\end{align*}

We now find the eigenvalues of the dynamics matrix $\mathbf{A}$, that determine the dominant dimensions of motion of the system: 
\begin{align*}
\lambda_1 =& - k_{1} + w_{1}, \; 
\lambda_2 = - k_{2} + w_{2}, \; 
\lambda_{3,4} = - \frac{k_1+k_2+w_1+w_2}{2} 
\pm  \frac{1}{2} \sqrt{16 \nu^{2} + \left(k_{1} - k_{2} + w_{1} - w_{2}\right)^{2}}. 
\end{align*}

If the cross-discriminator couplings $\nu$ are much smaller than the within-discriminator couplings $k_1, k_2, w_1, w_2$, then $\lambda_3$ and $\lambda_4$ are much larger than $\lambda_1$ and $\lambda_2$, and the system quickly converges to a plane attractor. In this case, the eigenvectors associated with $\lambda_1$ and $\lambda_2$ are $v_1^\top = [1\ -1 \ 0 \ 0]$ and $v_2^\top = [0 \ 0\ 1\ -1]$, respectively. Let $V= [v_1 \; v_2]$ and $\vec{x} = V \vec{y}$ such that $x_1 = y_1-y_2$ and $x_2 = y_3 -y_4$. Then, left-multiplying~\eqref{eq:4d-lca} by $V^\top$ yields the following slower dynamics:
\begin{align} \label{eqn:reduced_4d}
  \d \vec{x}= \vec{\tilde{r}} + \mathbf{\tilde{A}} \vec{x}   + \mathbf{S} \d \tilde{\mathcal{W}}_t, 
  \end{align}
where 
\[
\vec{r}= \begin{bmatrix} r_{1} - r_{2}\\
  r_{3} - r_{4}\end{bmatrix},\; 
 \mathbf{\tilde{A}}= 
  \left[\begin{matrix}- k_{1} + w_{1} & 0\\0 & - k_{2} + w_{2}\end{matrix}\right], \; 
\mathbf{S}\mathbf{S}^{\top}=
\begin{bmatrix}
\sigma_{1}^{2} + \sigma_{2}^{2}
- 2 \sigma_{12}   &  \sigma_{13} + \sigma_{24} -  \sigma_{14}  - \sigma_{23}  \\\sigma_{13}  + \sigma_{24} -  \sigma_{14}  - \sigma_{23}
& \sigma_{3}^{2} + \sigma_{4}^{2} - 2 \sigma_{34} \end{bmatrix}. 
\]

A few things are worth noting about expression~\eqref{eqn:reduced_4d}. First, the deterministic dynamics behave exactly as the 1-D diffusion model: Each of the two discriminators $x_1$ and $x_2$ accumulates differences among the stimulus-sensitive unit activities (i.e., $y_1-y_2$ and $y_3-y_4$). Either decay or self-excitation is possible, depending on the parameterization of the higher-dimensional system; specifically, on the signs of $(k_1-w_1)$ and $(k_2-w_2)$. Second, noise correlations \emph{within} discriminators in the 4D system ($\sigma_{12}$ and $\sigma_{34}$) are realized as noise \emph{variance} in the discriminator of the 2D system, whereas noise correlations \emph{across} discriminators in the 4D system are realized as noise \emph{correlations} in the 2D system. Taken in sum, this reduction yields a system exactly isomorphic to the probabilistic derivation above, providing convergent motivation for our model from probabilistic and mechanistic grounds. Finally, an additional response read-out layer can yield arbitrary response transformations of these stimulus discriminators using an elementary neural network \cite{Hornik1991}.

\section{Modeling Specific Tasks}

A variety of tasks can be modeled in the proposed framework by changing the initial condition distribution and response rule over the two-dimensional space, including two-alternative choices like such as AX-CPT and Flanker, three-alternative choices such as the prospective memory paradigm, and four-alternative choices that require identification of the precise identity of both stimuli. In the last case, a fixed threshold policy over the max-a-posteriori stimulus pair is the asymptotically-optimal MSPRT \cite{Baum1994,Dragalin1999}) previously used to model multi-hypothesis decisions in a number of domains \cite[e.g.][]{Shvartsman2014a,Norris2006,Bicknell2010Regressions,Bogacz2007NeuralComputation,McMillen2006}. Thus, the proposed framework provides a consistent way to view and analyze tasks that operate on two stimuli but vary in stimulus presentation timings and response mappings. We illustrate this using the Flanker and AX-CPT tasks as examples. 

\subsection{The Flanker task}

We begin by developing a model of the Flanker task \cite{Eriksen1974}, which can be cast in our framework as a task in which the response depends only on the target stimulus, and contextual information influences the decision only through its contribution to the inferred posterior over the target.
In this task, participants are shown a central target (e.g.\ `>' or `<') surrounded on both sides by distractors (`flankers') that are either congruent or incongruent with the target, and that we treat as the context in this paradigm. Participants are told to respond only to the target, but typically exhibit indications that processing is influenced by the context of the flankers, including an early period of below-chance performance and slower and/or less accurate responses when flankers are incongruent with the target relative to when they are congruent with it \cite{Gratton1988}. We label the two possible target identities $\{g_0$=`<',$g_1=$`>'$\}$ and the possible flanker identities $\{c_0=$<\_<,$c_1=$>\_>$\}$ with the underscore representing the position of the target. This gives us the two congruent possibilities $\{[C=c_0,G=g_0],[C=c_1,G=g_1]\}$ or [$<<<$,$>>>$] and the two incongruent possibilities $\{[C=c_0,G=g_1],[C=c_1,G=g_0]\}$ or [<><,><>]. The response mapping is independent of the context (i.e., the flankers) and is given by 
\begin{align*}
r(c, g)=
\begin{cases}
r_0, &\mbox{ if } g= g_0,\\
r_1, &\mbox{ if } g=g_1.  
\end{cases}
\end{align*}
Thus, the set of stimuli that map to $r_0$ and $r_1$ are $\Gamma_0 = \{\{c_0, g_0\}, \{c_1, g_0\}\}$ and $\Gamma_1 = \{\{c_0, g_1\}, \{c_1, g_1\}\}$, respectively. It should be noted that our model~\eqref{eqn:2d-diffusion-decay} is a continuous-time limit of an earlier model of the Flanker task \cite{Yu2009}, capturing that model as a special case of our formulation.

\subsubsection{Fixed threshold policies for the Flanker task} 

Recall that the fixed threshold policy compares the log-likelihood ratio of a response being correct $Z(t)$ defined in equation~\eqref{eq:response-LLR-continuum} with two thresholds. Tailoring the expression in equations~\eqref{eq:response-LLR} and ~\eqref{eq:response-LLR-continuum} to the Flanker task, we obtain
\begin{align}
Z(t) &=  z_g(t) + \log\frac{P_{11} e^{z_c(t)}+ P_{01}}{P_{10}e^{z_c(t)} +P_{00}}, 
\label{eqn:Flanker}
\end{align}
where $P_{ij}= P(C= c_i, G= g_j)$. The statistic $Z(t)$ in equation~\eqref{eqn:Flanker} comprises the DDM term for the target $z_g(t)$ biased by the nonlinear function of the DDM term for the context $z_c(t)$. We refer to this bias as the \emph{dynamic context} term. 

We consider the dynamic context term when the prior on the target and context are independent, $P_{ij}= P(C= c_i, G= g_j)= P(C= c_i)P(G= g_j) = p_i q_j$. In this case, 
\[
\log\frac{p_1 q_1 e^{z_c(t)}+ P_{01}}{P_{10}e^{z_c(t)} +P_{00}} = \log\frac{p_1 q_1 e^{z_c(t)}+ (1-p_1)q_1}{p_1 (1-q_1)e^{z_c(t)} +(1-p_1)(1-q_1)} = \log \frac{q_1}{1-q_1},
\]
which is the prior log-likelihood ratio for the target. Therefore, in this case, the Flanker update~\eqref{eqn:Flanker} reduces to the standard DDM.

For simplicity, we choose the thresholds for the fixed threshold policy~\eqref{eq:ftr-rule} symmetrically, i.e., $A= -B =\theta$. Accordingly, at the time of a response we have
\[
z_g(t) + \log\frac{P_{11} e^{z_c(t)}+ P_{01}}{P_{10}e^{z_c(t)} +P_{00}}= \pm \theta,
\]
which can be interpreted as a set of curves in the two-dimensional $\vec{z}$ space.

\begin{figure}[ht!]
    \centering    \includegraphics[width=0.8\linewidth]{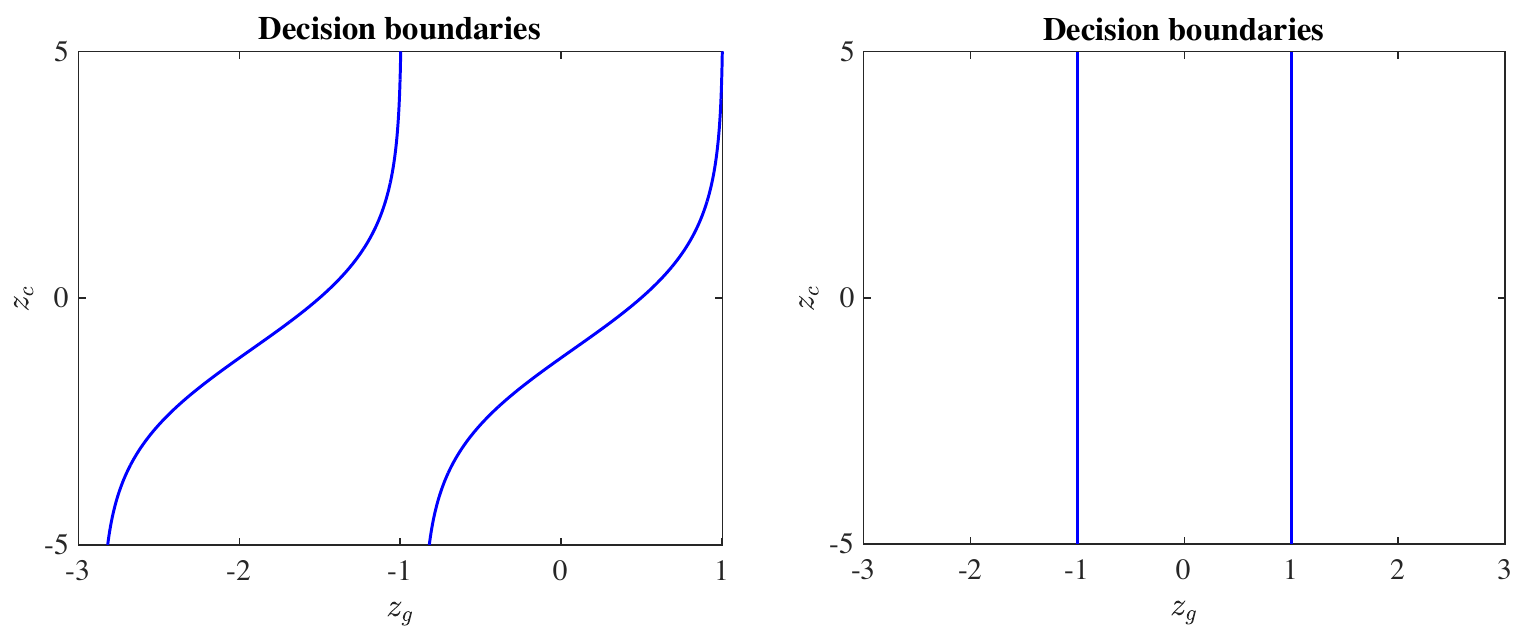}
    \caption{Decision boundaries for the Flanker task in $\vec{z}$ space. When the priors on the context and the target are correlated, the evidence from the context $z_c$ determines the required evidence for the target $z_g$ (left panel). When the prior probabilities are uncorrelated, then context plays no role in the decision and the model reduces to the standard DDM with fixed boundaries (right panel).  } %
    \label{fig:flanker-boundaries}
\end{figure}

When the prior joint distribution over context and target reflects statistical dependence between them, the context belief contributes to the statistic $Z(t)$ and dynamically biases the amount of target evidence required for making a response, as shown in Figure~\ref{fig:flanker-boundaries}. This influence arises through the prior over the context-target pair, even though the sensory samples from context and target are assumed independent. In contrast, when the prior factorizes, so that context and target are independent under the observer's generative model, the context provides no information about the correct response, and the model reduces to the standard DDM. In this formulation of the policy (fixed-threshold), the observed flanker effect emerges naturally from this prior dependence together with the corresponding drift directions of context and target evidence. We discuss in Section~\ref{sec:bro-flanker} alternate explanations for the flanker congruence effect under the CDDM framework.

\paragraph{Comparison to a constant-drift model}
\label{sec:Flanker-ddm}

The expressions above in terms of likelihood ratios facilitate comparison to Wald's SPRT and Wiener diffusion models, such as the DDM. This is complementary to an earlier approach performing dynamical analysis on the problem in probability space \cite{Liu2009}. Since the likelihood ratio and the a posteriori probabilities are monotonically related, thresholding on $Z(t)$ maps onto the threshold over the probability of the most probable response described above. $Z(t)$ in equation~\eqref{eqn:Flanker} comprises two terms. The first is the unbiased SPRT statistic on the target, while the second is a nonlinear function of the SPRT statistic for the decision on the context. In the case where the context is predictive of the target, the model has a \emph{congruence bias} wherein the nonlinear term plays the role of bias in the SPRT for decision on target. This rational dynamic prior bias is an advance over previous heuristic approaches to dynamic biases~\cite[e.g.][]{Hanks2011}. As will be shown in Section~\ref{sec:bro-flanker}, this is one of three distinct explanations for flanker congruence effects in our model. 

\begin{figure*}
\includegraphics[width=\textwidth]{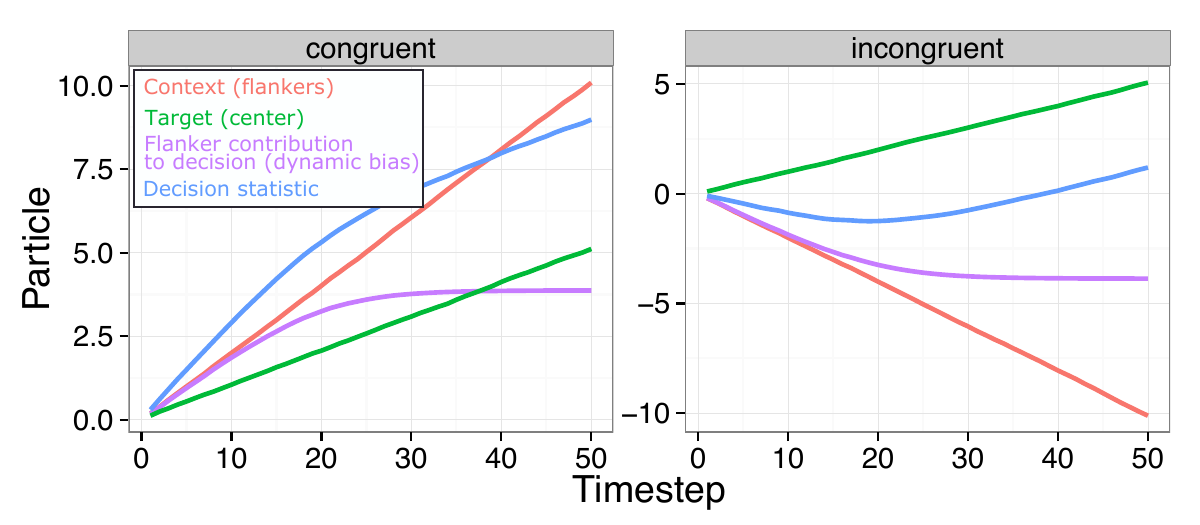}
\caption{\textbf{Average trajectories in the external context model with congruence bias on a congruent (left) and incongruent (right) trial.} This model is equivalent to the congruence-bias model of the Flanker task \cite{Yu2009}, in which statistical dependence between flankers and target produces dynamic biases in the decision variable on both congruent and incongruent trials, though it can just as well be applied to similar paradigms such as the Stroop task. In terms of log-likelihoods, the external context model decomposes into a linear log-likelihood random walk on the target and a saturating bias term. On a congruent trial, the influence of the flankers accelerates the progress of the decision variable. On an incongruent trial, the decision variable moves away from the true response early on due to the initial influence of the context, but eventually reverses under the influence of the target. }
\label{fig:Flanker-decomposed}
\end{figure*}

\begin{figure*}
\includegraphics[width=\textwidth]{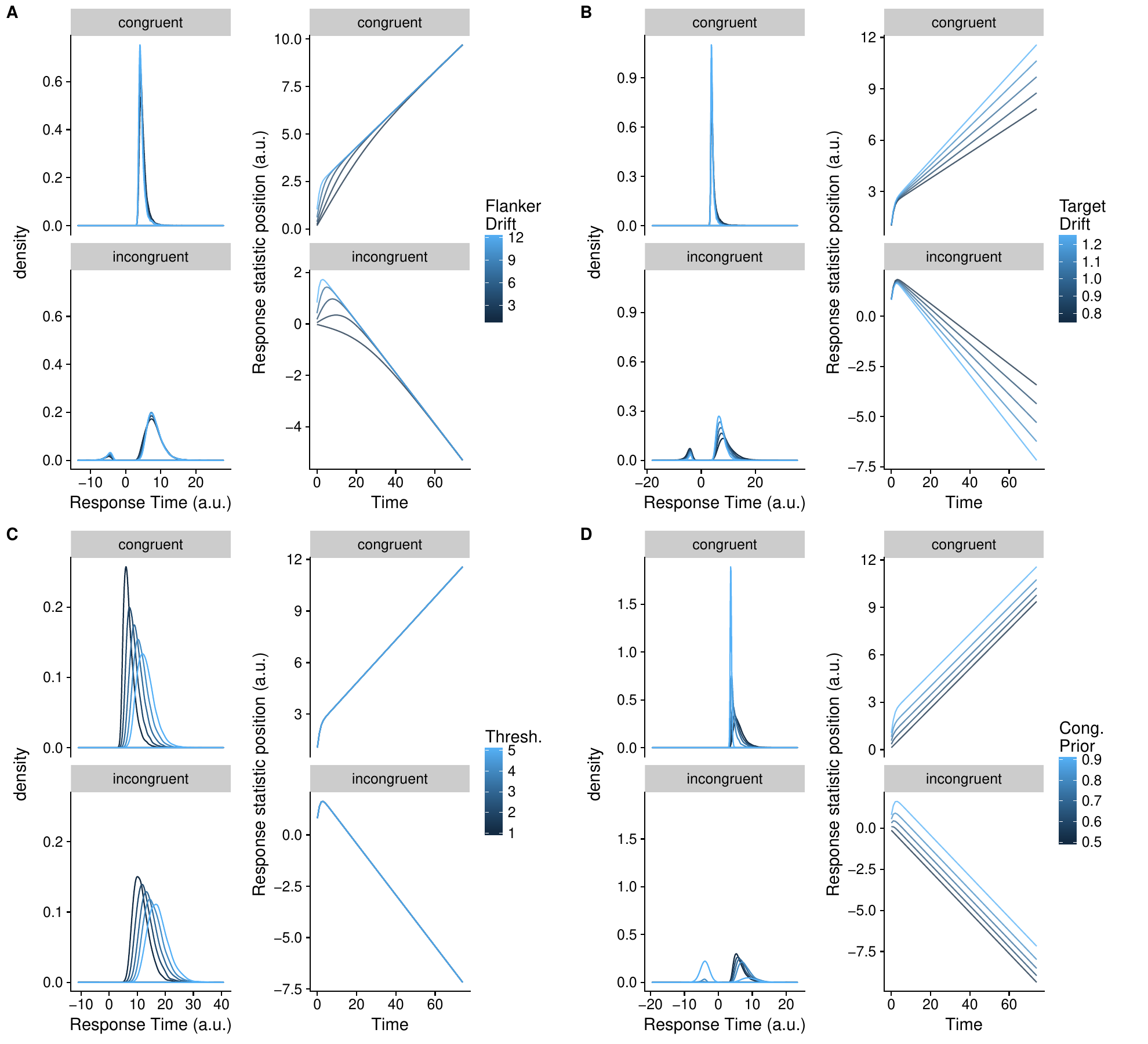}
\caption{The decision variable in the model as the prior over congruence, context drift, and target drift are varied. Congruence affects the timing and proportion of early errors nonlinearly and later RTs linearly; context drift affects the timing and proportion of early errors linearly and later RTs not at all (once the context bias is overcome); target drift affects later RTs linearly (as in the fixed context model).}
\label{fig:Flanker-parsweep}
\end{figure*}

Figure~\ref{fig:Flanker-decomposed} illustrates this decomposition, by showing the nonlinear random walk for this task decomposed into the linear term for the central stimulus and a saturating term for the flankers. In this decomposition, we can see the independent contributions of the drift on the context and target, as well as the a priori bias, to the final decision and therefore the predicted RT distributions. In particular, the target drift scales the random walk linearly, the context drift scales it only on early RTs (since later their effects cancel out), and the a priori bias has a nonlinear effect both early and late (see Fig.~\ref{fig:Flanker-parsweep} for further illustration of this point).

This model is isomorphic to the Flanker model of \cite{Yu2009}, in which statistical dependence between flankers and target produces dynamic biases in the decision variable on both congruent and incongruent trials.

The model reduces to a fixed context DDM in reasonable cases. As discussed earlier, when the priors on the context and target are uncorrelated, 
i.e., the context is uninformative, the model reduces to DDM. 
Likewise, when the prior joint distribution over context and target reflects perfect positive or negative dependence, e.g., when $P_{01}=P_{10}=0$ or $P_{11}=P_{00}=0$, the model reduces to the sum or the difference of the context and target DDMs, respectively, corresponding to congruent and incongruent conditions.
Also, when $|z^c_{\tau}|$ is large (i.e.\ the context is perfectly known), the model approaches a biased fixed-context model. If the context prior is informative, then on early timesteps either the numerator or the denominator of the dynamic bias expression is pushed in the direction of the prior (due to the conditional probability term). Still, as the context statistic grows, it dominates the numerator and denominator of the dynamic bias term. The overall term approaches 0 -- in this case, the context is perfectly known and we are again in the fixed context DDM case.
Finally, if the magnitude of the drift rate for the context is much higher than the magnitude of the drift rate for the target, or the magnitude of the bias $z_0^c$ is high, then the nonlinear term saturates at a faster timescale compared to the decision time. In this limit, the approximate contribution of the nonlinear term is either $\log\left(\frac{P_{01}}{P_{00}}\right)$,
or  $\log\left(\frac{P_{11}}{P_{10}}\right)$ and once again the model reduces to the DDM.

\subsubsection{Bayes risk optimal policies for the Flanker task}
\label{sec:bro-flanker}

We compute the BRO policy using the framework discussed in Section~\ref{sec:decision policies}. 
Parameter values were chosen to reflect qualitative features of the Flanker task while avoiding degenerate regimes in which one stimulus dominates the decision. In particular, we set the context and target drift rates to 0.3 and 0.1, respectively, reflecting the greater perceptual salience and signal strength of flankers relative to the target.  To assess robustness, we explored a broad range of parameter values, varying drift rates, prior dependence, and sampling costs over approximately an order of magnitude around these nominal settings. Across this range, the qualitative structure of the optimal policy and the resulting behavioral signatures remained unchanged, with differences appearing only in overall response times and error rates. The sampling costs were set to $L_c=L_g=1$, $\subscr{L}{both}=1.5$, and $\subscr{L}{incrt}=40$. These values reflect the assumption that sampling incurs both opportunity and effort costs, so that sampling both stimuli simultaneously is more costly than sampling either alone, but less costly than the sum of the individual costs due to shared processing resources.

To solve for the optimal policy, we discretize the continuous belief space on a grid, which makes the transition densities for the sampling actions mixtures of (discretized) Gaussian distributions with means $\pm\vec{a} - \Lambda\vec{z}$ and identity covariance. Specifically, we discretize the belief space to a $110\times110$ grid between $\pm3$ in both dimensions and solve it by backward induction in discrete time, with time discretized at $\d t=1$, and the horizon set well beyond typical decision times. The solutions we show apply to all possible priors, which only determine where on the grid the decision variable starts.

\begin{figure*}[ht!]
  \begin{center}
    \includegraphics[width=0.3\textwidth]{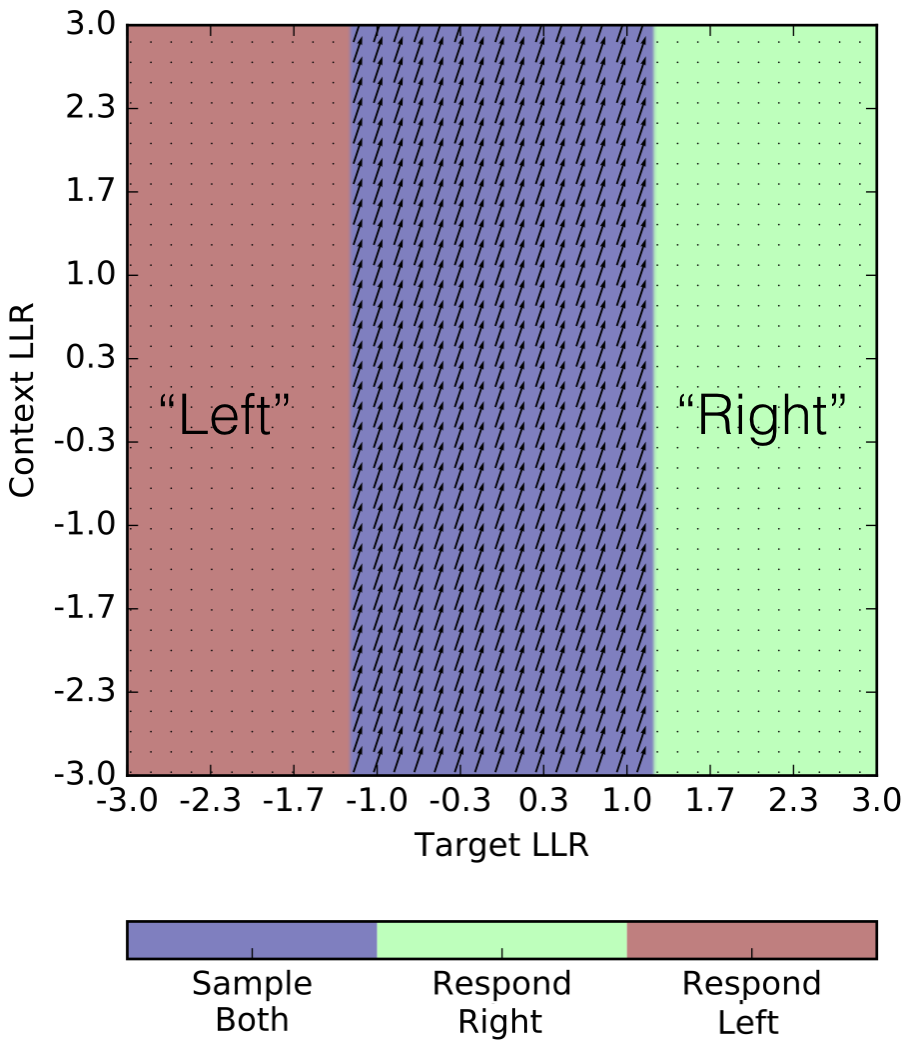}
    \includegraphics[width=0.3\textwidth]{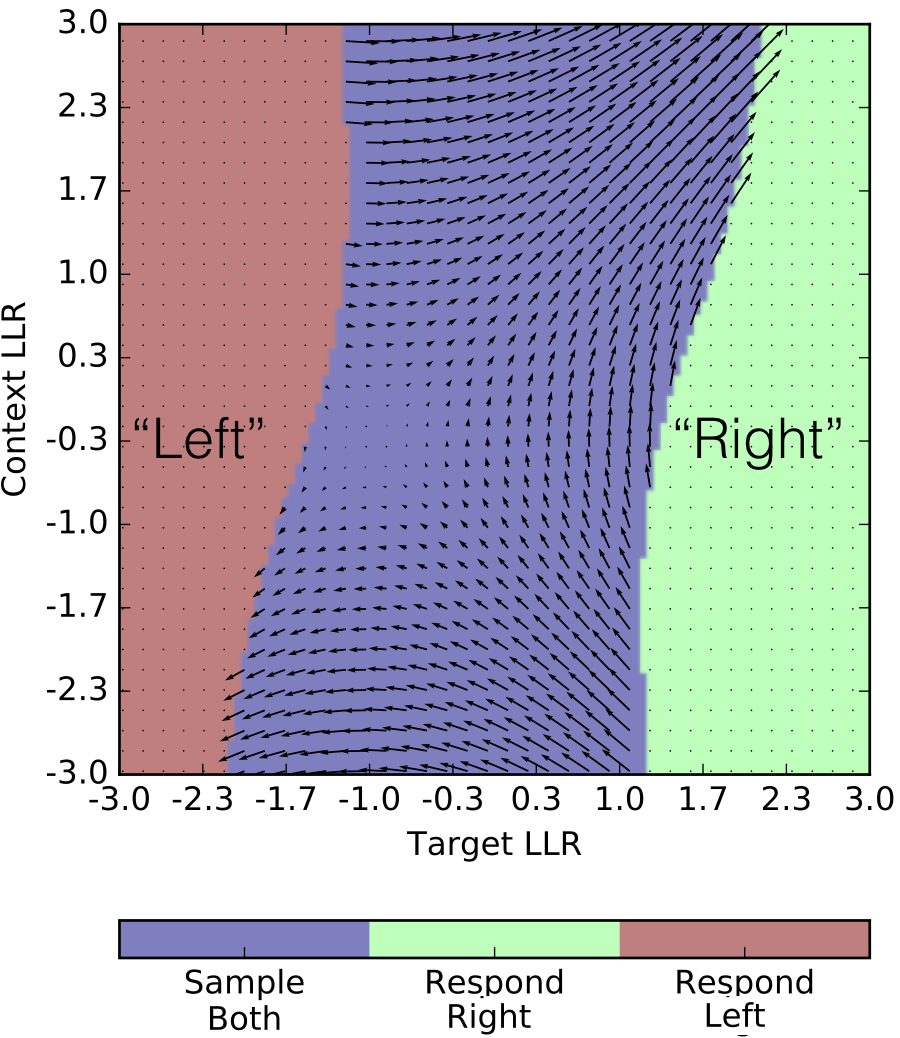}
      \includegraphics[width=0.3\textwidth]{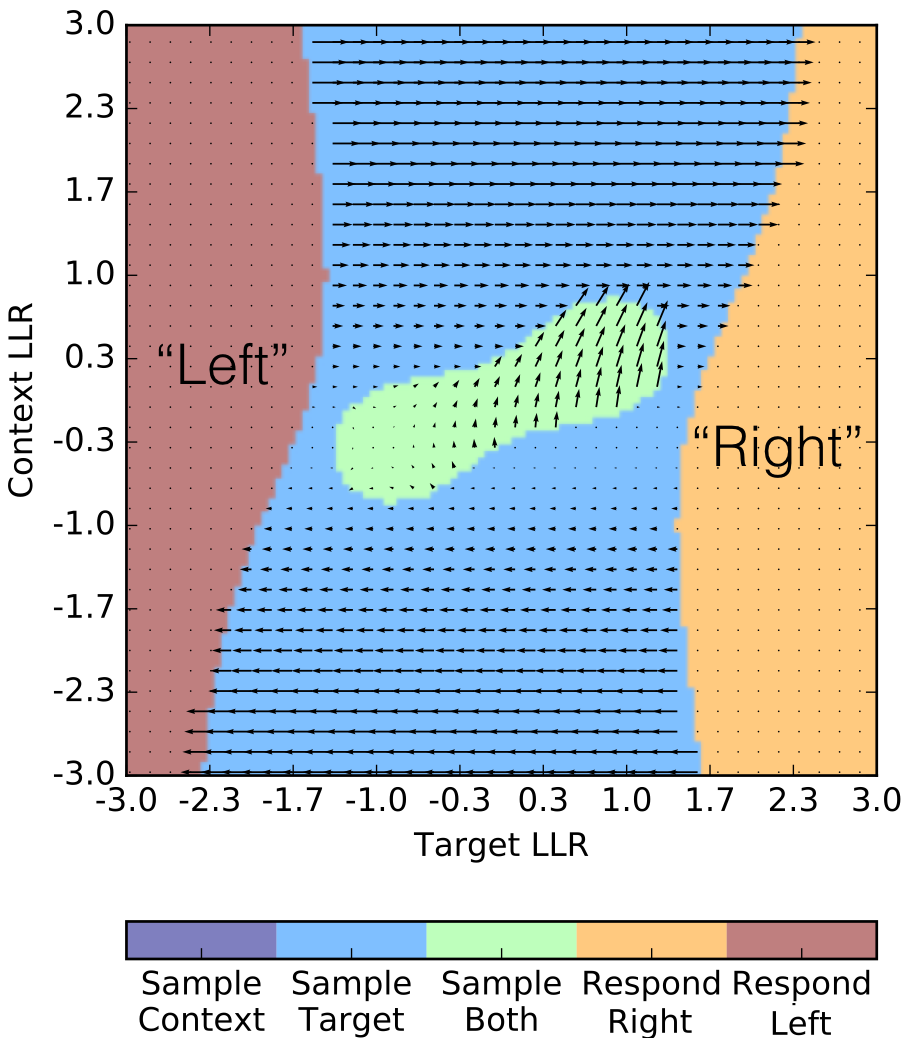}
  \end{center}
  \caption{\textbf{Optimal policies for the Flanker task}. \emph{Left}: Optimal policy without spatial uncertainty and obligatory sampling of both stimuli. \emph{Middle}: Optimal policy for the same model with spatial uncertainty added, producing characteristic biases similar to those described in Flanker models such as \cite{Yu2009,White2011}. \emph{Right}: Same model with the ability to sample stimuli independently. Vector fields illustrate the drift direction of $\vec{z}(t)$.}
  \label{fig:Flanker-opt}
\end{figure*}

Participants are slower and less accurate on incongruent Flanker task trials and are below chance accuracy on fast responses \cite{Gratton1988}. This effect has been explained in sequential sampling models by making one of three different assumptions: (a) that 
participants have a bias to expect congruent trials \cite{Yu2009}; (b) that perceptual uncertainty makes samples from target and Flankers correlated \cite{Yu2009}; and (c) that a shrinking attentional spotlight admits Flanker samples into decisions early but not late \cite{White2011}. We show how all three solutions can be optimal under different assumptions about human cognitive limitations. 

\
\paragraph{Fixed-threshold model (with correlated prior)}
We first consider a model with no spatial uncertainty and no ability to selectively sample stimuli. In this case, as shown in Fig.~\ref{fig:Flanker-opt}~(left), the optimal BRO policy reduces to a fixed-threshold policy on $z_g$ that depends on the context belief $z_c$. 
Any apparent congruency effects can arise only through differences in the initial condition of $z_g$. For example, if congruent trials are assumed to begin with an initial bias toward the correct response (positive or negative $z_g$), the decision variable will reach the threshold more quickly, resulting in faster and more accurate responses. Conversely, if incongruent trials begin with an initial bias toward the incorrect response, they will exhibit slower responses and a greater likelihood of early errors. However, these effects arise from the assumed initial condition of $z_g$ rather than from any principled use of contextual information in the inference or stopping rule. Intuitively, the initial condition of $z_g$ lies on a curve crossing Fig.~\ref{fig:Flanker-opt}~(left) diagonally, and Fig.~\ref{fig:flanker-boundaries}~(left) absorbs this prior effect into the thresholds.

\paragraph{Perceptual uncertainty model} The second model implements spatial uncertainty by setting the off-diagonal terms of $\Lambda$ to $-0.3$, introducing mutual excitation between the stimulus accumulators. As shown in Fig.~\ref{fig:Flanker-opt}~(middle), the resulting BRO policy resembles a fixed-threshold policy with correlated priors on the context and target, as shown in the left panel of Fig.~\ref{fig:flanker-boundaries}.  Here, decision boundaries for $z_g$ vary with $z_c$.
In the congruent quadrants, where context and target evidence tend to support the same response, new samples move the belief more rapidly toward a decision, increasing the value of continued sampling. As a result, the optimal policy delays commitment in these regions by placing the decision boundaries farther away, which can counteract the tendency toward faster responses on congruent trials. Conversely, in the incongruent quadrants, where context and target evidence favor opposing responses, the belief evolves more slowly, reducing the value of continued sampling and leading the optimal policy to commit earlier. These two effects are in opposition, with the increased drift increasing the size of the congruence effect and the rising threshold reducing it. For this perceptual uncertainty model to yield the standard flanker congruence effect under a BRO policy, the impact of the drift increase has to be greater than the impact of the threshold increase.

\paragraph{Attentional spotlight model} The third model combines spatial uncertainty with the ability to selectively sample either stimulus. As shown in Fig.~\ref{fig:Flanker-opt}~(right), the optimal policy initially samples both stimuli when uncertainty is high, and then shifts sampling toward the target as the context becomes more certain. The transition depends on the statistical dependence between context and target: when context samples substantially reduce uncertainty about the target, continued sampling of the Flankers remains valuable over a larger region of the belief space, whereas when they do not, the policy shifts attention to the target sooner. In this way, the model exhibits a rudimentary adaptively shrinking attentional spotlight \cite{White2011}.

This behavior is consistent with normative models of fixation allocation and active sensing, in which attention and sampling choices are part of a unified inference process that optimally balances information gain with costs~\cite{callaway2021fixation,jang2021optimal}. Although the spotlight shrinks discretely in stimulus space, its interaction with compatibility bias can produce the appearance of gradual shrinking in response space.

Taken together, these models show that the major explanations for Flanker effects can be understood as normative consequences of different policy constraints and action sets, consistent with bounded-optimal or computationally rational accounts \cite{Lewis2014}. The parameter settings shown here are illustrative; intermediate regimes arise under other parameter choices.

\subsubsection{Discussion}

In this section, we have demonstrated how a model of the Flanker task can be implemented in the theoretical framework we have presented. This illustrates how this framework can bring together a number of different results. From a descriptive perspective, the model is a continuous-time variant of the model of \cite{Yu2009}, which allows us to provide a continuous-time analog of the analyses of \cite{Liu2009}. From an explanatory perspective, the framework permits the derivation of a normative dynamic bias in this task (in contrast to the heuristic dynamic bias of \cite{Hanks2011}), and demonstrates how the explanations for the Flanker effect of both \cite{Yu2009} and \cite{White2011} can be viewed as optimal based on different assumptions about human computational bounds. 

\subsection{The AX-CPT}

\begin{figure}[ht!]
  \centering
  \includegraphics[width=0.7\columnwidth]{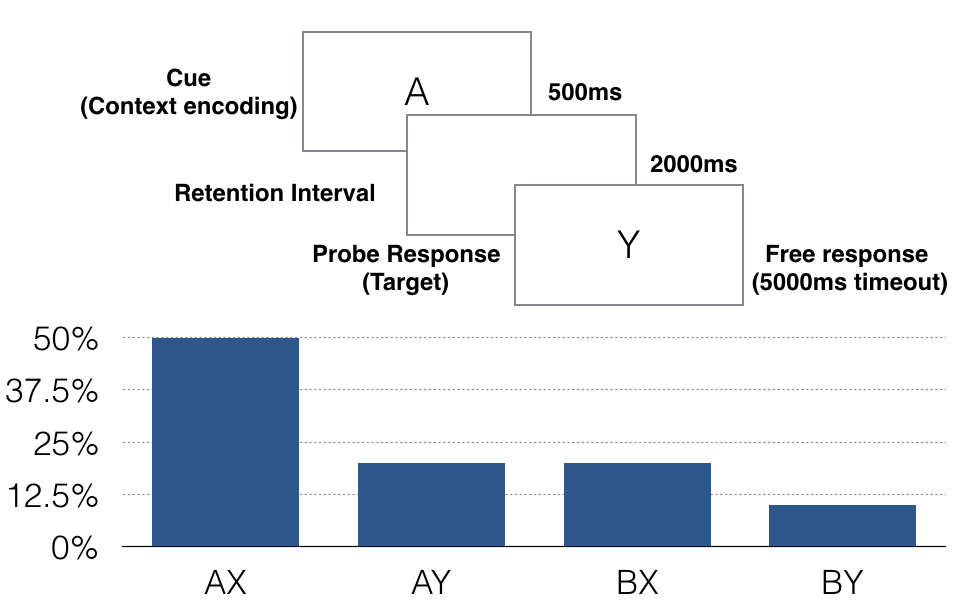}
  \caption{\emph{Top}: sequence diagram for one trial of the AX Continuous Performance Test. \emph{Bottom}: frequency of each trial type in the experiment. %
  }
  \label{fig:axcpt-task}
\end{figure}

In contrast to the Flanker task, which is an \emph{external} context task with a context-\emph{independent} response, the AX-CPT is an \emph{internal} context task with a context-\emph{dependent} response. 
In the AX-CPT, participants are asked to make a response to a probe (target) stimulus, by convention labeled `X' or `Y', for which the response mapping is determined by a previously seen cue (context) stimulus, `A' or `B'. In our notation: $g_0=X,g_1=Y,c_0=A,c_1=B$. Unlike the Flanker, for which all stimuli pairs are usually equally likely, in the AX-CPT, AX trials are usually the most common (appearing 50\% of the time or more), and BY trials are the least common. AY and BX trials appear with equal frequency but have different conditional probabilities due to the preponderance of AX trials. 
Note that, for the case in which AX and BY trials are assigned one response, and AY and BX trials are assigned the other response (as in the treatment here), the mappings correspond to an exclusive-or operation. However, this need not always be the case \citep[e.g.][]{Servan-Schreiber1996}. A schematic diagram of the task is shown in Figure~\ref{fig:axcpt-task}.

\subsubsection{Fixed Threshold Policy for the AX-CPT}

To compute the fixed threshold policy for the AX-CPT, we first specialize the expression~\eqref{eq:response-LLR-continuum} with $\Gamma_1=\{\{c_0, g_0\},\{c_1, g_1\}\}$ and $\Gamma_0=\{\{c_0, g_1\},\{c_1, g_0\}\}$ to the obtain the log-likelihood of $r_1$ being the correct response as
\begin{align}\label{eqn:axcpt-sym}
Z(t)= \log \frac{P_{00}e^{z_c(t)}e^{z_g(t)}+P_{11}}{P_{01}e^{z_c(t)}+P_{10}e^{z_g(t)}},
\end{align}
where $P_{ij}:= P(C=c_i, G=g_j)$ are the prior probabilities on the context and target pairs. 

Similarly to the Flanker task, we select the thresholds symmetrically, i.e., $A=-B=\theta$. Thus, at the time of the decision, the evidence $z_c$ and $z_g$ satisfy
\[
\log \frac{P_{00}e^{z_c}e^{z_g}+P_{11}}{P_{01}e^{z_c}+P_{10}e^{z_g}} = \pm \theta \iff z_g =  \log\frac{ - P_{11}+ P_{01}e^{z_c}e^{\pm \theta}}  {P_{00}e^{z_c}- P_{10}e^{\pm \theta}} \iff z_c =  \log\frac{- P_{11}+ P_{10}e^{z_g}e^{\pm \theta}}{P_{00}e^{z_g}-  P_{01}e^{\pm \theta}}.
\]

\begin{figure*}
\centering 
\includegraphics[width=0.5\textwidth]{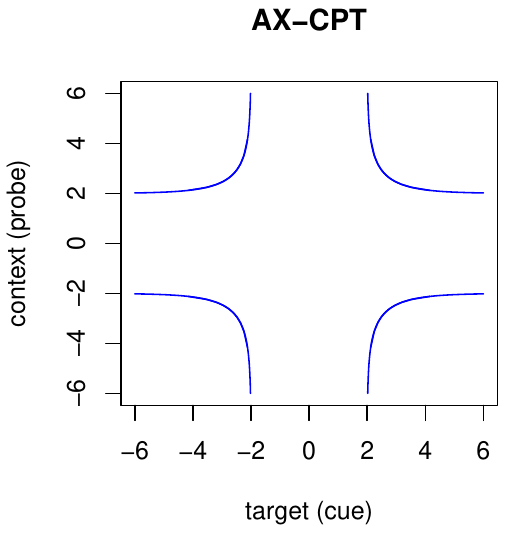}
\caption{\textbf{Shape of fixed threshold over the response in stimulus belief space for AX-CPT.} Both thresholds have the same value when defined over the response statistic $Z$. However, when mapping them back into the 2D space of the context and target statistics $z_c$ and $z_g$, the thresholds lead to curved boundaries determined by the AX-CPT response rules and frequencies.} 
\label{fig:cddm-2d}
\end{figure*}

As with the Flanker, these equations yield a set of boundaries in the two-dimensional space of $\vec{z}$. 
Figure~\ref{fig:cddm-2d} shows the shape of the boundary in 2D space for AX-CPT. The boundary asymptotes as $e^{z_c} \rightarrow \frac{P_{11}}{e^{\pm \theta} P_{01}}$ and $e^{z_c} \rightarrow \frac{e^{\pm \theta} P_{10}}{P_{00}}$, which is the evidence required on the context to make a decision when the target is perfectly known. Crucially, when the target is perfectly known, the decision dynamics reduce to a DDM for the context.  A similar asymptote exists corresponding to the case of context being perfectly known, which maps to a DDM for the target.

As in the case of the Flanker model, each parameter in the AX-CPT model has both a distinct psychological interpretation and distinct effects on response time distributions.  The threshold reflects the speed-accuracy tradeoff, and therefore affects the shape of the RT distributions and the error proportion across all trial types. The three context parameters affect different portions of the RT distribution: the context encoding drift contributes primarily to the early portion of the RT distributions, due to how it affects the initial condition for the decision; decay reflects memory retention, and contributes both to early and late portions of RT distributions because it persists both during retention and retrieval; and context retrieval drift reflects memory retrieval, and contributes primarily to middle and late portions of the RT distributions because retrieval begins at target onset and saturates over time. All three context parameters identically affect trial types that share a context (e.g.\ AX and AY). In contrast, target drift reflects perceptual processing of the target stimulus and therefore affects the RT distribution throughout, but identically affects the same-target pairs (e.g.\ AX and BX).

Thus, differences between trial types reveal the relative contributions of context and target processing, while differences across portions of the RT distribution within a trial type reflect distinct subcomponents of context processing, including encoding, retention, and retrieval.
For example, high $a_e$ increases the number of very fast correct responses when the context predicts the correct target (e.g.\ AX), but also contributes fast errors when it does not (e.g.\ AY). In this way, the model provides a more precise characterization of context processing differences than do behavioral indices alone \cite[e.g.][]{Braver2012}, which may be insufficient to distinguish perceptual and memory-based processing \cite{Lositsky2015}. 

\begin{figure*}[ht!]
\includegraphics[width=\textwidth]{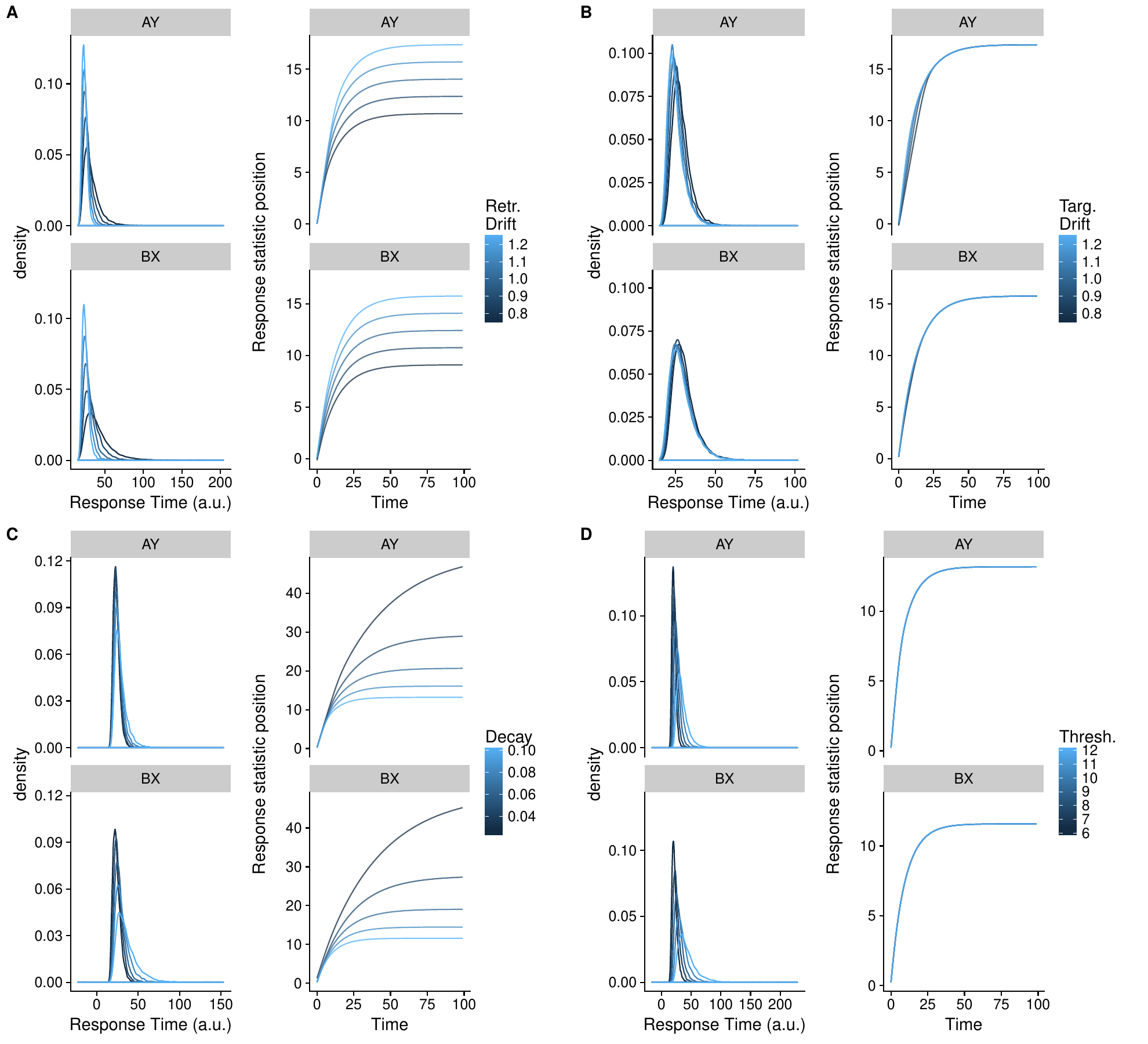}
\caption{\textbf{Average and response time distributions in the internal context (AX-CPT) model, showing different effects of different parameters.} Effects are shown for AY and BX trial types, which are illustrative of the effects of primary interest (figures for the AX and BY are in the appendix). Note that the threshold and retrieval parameters affect AY and BX trials differently, with threshold shifting the mode and retrieval drift shifting the tail. In contrast, target drift has its affect primarily on AY trials (as this determines the extent to which responses in this condition would be confused with AX trials), whereas decay has its effect primarily on BX trials (by changing the memory of the context).}
\label{fig:axcpt-parsweep}
\end{figure*}

\paragraph{Comparison to a constant-drift model} 
In contrast to expression~\eqref{eqn:Flanker} for the Flanker model, there is no straightforward way to extract a linear drift on either context or target in the AX-CPT model. Accordingly, to facilitate comparison to the conventional DDM, we focus on the response statistic for 1D nonlinear diffusion model~\eqref{eqn:axcpt-sym}.
Equation~\eqref{eqn:axcpt-sym} combines the DDM associated with the context and the target in a nonlinear fashion, and the reduction of these equations to a DDM or the sum of two DDMs is valid only in restricted settings. For example, if the drift rate for the target is much higher than the drift rate for the context, then in the limit of large thresholds~\eqref{eqn:axcpt-sym} can be approximated by either $\log \frac{P_{00}}{P_{10}} + z_c(t)$, or $\log \frac{P_{11}}{P_{01}} -z_c(t)$. A similar one-dimensional approximation is possible if the drift rate for the context is much higher than the drift for the target.

\subsubsection{Bayes Risk Optimal Policy for AX-CPT} 

As with the Flanker model, we can find the BRO sampling and response policies for the symmetric AX-CPT model, 
using the POMDP framework discussed in~\ref{sec:decision policies} to determine the optimal policy for AX-CPT. 
The parameters used in the simulation are $a_c=a_g=0.2$ and $\Lambda=0.1$. The discretization and other parameters are set as in the Flanker task. 
As in the Flanker analysis, the qualitative patterns are not strongly dependent on these parameters. 

\begin{figure}[ht!]
  \begin{center}
  \centering
    \includegraphics[width=0.5\columnwidth]{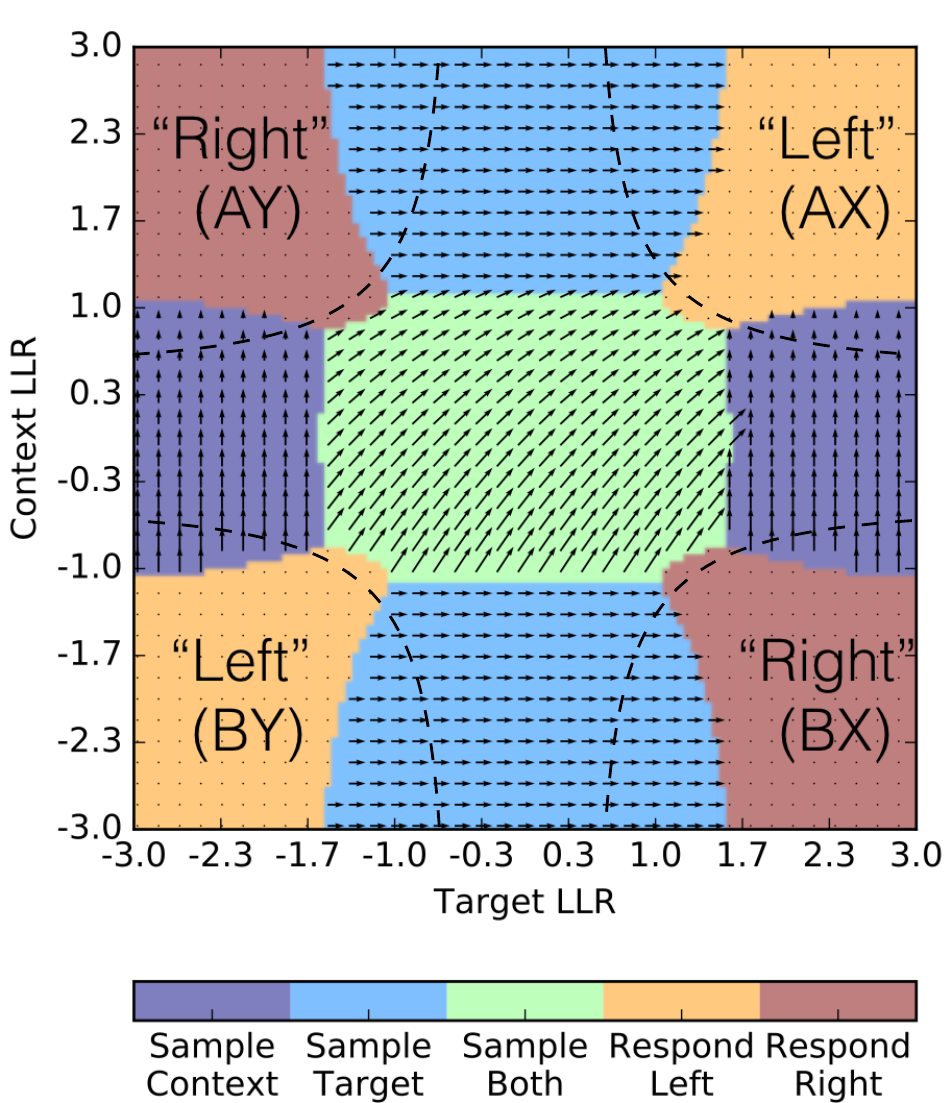}
  \end{center}
  \caption{\textbf{BRO policy for AX-CPT.} The dashed contour marks a possible FTR policy -- note the distinctively different shape relative to the BRO policy.
  }
  \label{fig:axcpt-opt}
\end{figure}

Figure~\ref{fig:axcpt-opt} (left) shows the optimal policy for this task. The BRO agent samples both stimuli until it has enough evidence for one, and then samples just the other. We observe that, in general, the BRO policy is not contained in the space of FTR policies. Rather, it makes the prediction that the decision bounds with respect to one stimulus widen as the posterior of the other concentrates. As with Flanker model, this is again because the value of sampling increases -- this time due to the structure of the task rather than interference. The intuition is that, with sufficient certainty that the first stimulus is `A', distinguishing between `X' and `Y' is valuable since that will allow a correct response. However, if there is uncertainty about the context stimulus, distinguishing between the target stimuli is less useful because responses based on one stimulus alone will have a high error probability (up to 50\% if all trial types are equally likely). This threshold increase also appears in the BRO (but not FTR) policy for the general multi-hypothesis setting \cite{Dragalin1999}. As such, these predicted effects may be useful for distinguishing between BRO and FTR policies. Furthermore, manipulating stimulus noise in the AX-CPT task may be useful for testing these predictions: a noisier signal of one stimulus should result in faster decisions. This is counterintuitive, as typically one would expect a more difficult stimulus to result in a slowing rather than speeding of responses. 

\subsubsection{Optimal cognitive control in AX-CPT} 

In addition to analyzing BRO policies, we can also analyze qualitative signatures of optimal attention allocation in this task. Because finding RR-optimal policies requires brute-force search in our setting, we simplify the problem by restricting analyses to the FTR decision policy
and search only over RR-optimal attention allocation policies.
The phenomenon of interest in this section is control allocation in the AX-CPT task. Cognitive control in such tasks is often considered within the Dual Mechanisms of Control (DMC) framework \cite{Braver2012}, which posits that control can be allocated in two ways: \emph{proactively} and \emph{reactively}. Proactive control requires effortful task \emph{preparation} and \emph{maintenance}, that can be modeled in our framework as a relatively large \emph{context encoding drift} and lower decay. In contrast, reactive control involves stimulus-driven \emph{retrieval} of a response rule (presumably from episodic memory), and can be modeled in our framework as a relatively large \emph{retrieval drift} and decay as well as a smaller context encoding drift. This ability to represent both strategies in the same model is in contrast to models that require qualitatively distinct machinery for both strategies \cite[e.g.][]{Lositsky2015}.

\begin{figure}[ht!]
\centering
\includegraphics[width=0.7\columnwidth]{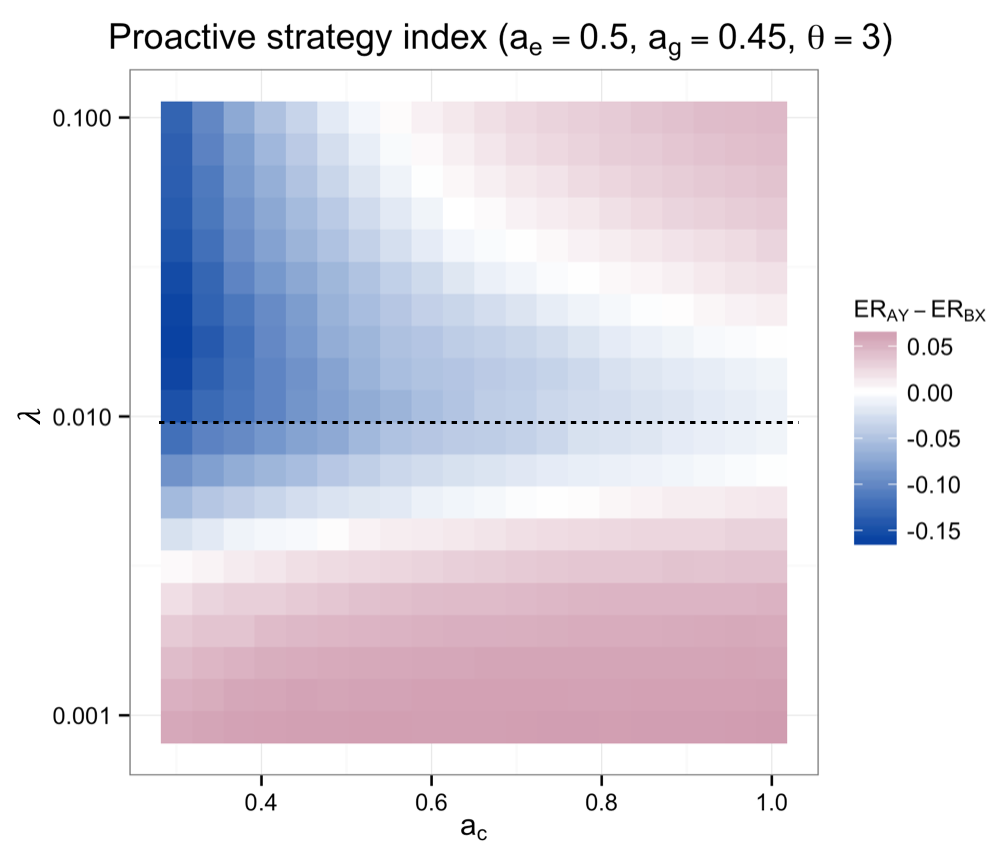}\\
\caption{Proactiveness index as a function of retrieval drift and memory decay. The dashed line marks the `complete forgetting' boundary, the magnitude of decay beyond which the context statistic decays to zero by the time retrieval starts. }
\label{fig:prore}
\end{figure}

Canonically, the difference in performance between AY and BX trials (which have the same marginal probabilities but different conditional ones) can be used to coarsely distinguish between proactive and reactive strategies. In particular, a proactiveness index can be defined as the difference in the number of \emph{correct} responses in AY trials and the number of \emph{incorrect} responses in BX trials. 
Figure~\ref{fig:prore} shows this proactiveness index for simulations of the AX-CPT using our model, computed with context encoding drift, target drift, and threshold held constant, and adjusting context retrieval drift and memory decay. The pink regions mark proactive strategies and the blue ones mark reactive ones. The ability to encode both strategies in one model allows us to predict two qualitatively different regimes of operation at different rates of memory decay ($\lambda$), separated by the dashed line in the figure. At lower $\lambda$, higher retrieval drift ($a_c$) is needed to maintain a proactive strategy as $\lambda$ increases. However, at higher decays, the context is forgotten completely and the pattern reverses: lower $a_c$ is needed to maintain a constant proactiveness index as $\lambda$ increases.

Both results are consistent across other parameters, with higher $a_g$ making the model more proactive. Furthermore, we can determine whether both patterns are normative by making assumptions about which parameters are cognitive bounds and which are under the participant's control, and optimizing behavior with respect to the latter. Specifically, we assume that the total sum of context and target drift is bounded, as is the memory decay, while the relative allocation of the context and target drift and the threshold are controllable. Assuming a fixed threshold policy allows us to use a reward rate~\cite{Bogacz2006} (averaged over 10,000 trials) rather than Bayes risk as the loss. This also avoids the need to perform dynamic programming to find the shape of the optimal threshold, which would be intractable when nested within the RR optimization; we can then solve for the optimal policy by gradient-free optimization \cite{Powell2009}. The benefit of this approach is that we can find the optimal policy for a set of 900 parametric architectures, each defined by a memory decay (log-spaced between 0.001 and 0.01 in 30 steps) and total drift (equally spaced between 0.5 and 4.5 in 30 steps), and thereby characterize the relationship between cognitive bounds and optimal control allocation. 

\begin{figure}[ht!]
\centering
\includegraphics[width=0.7\columnwidth]{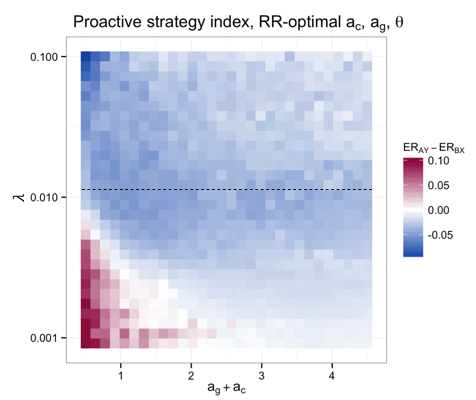}
\caption{Optimal proactiveness index as a function of total drift (retrieval + target) and memory decay. In contrast to Fig.~\ref{fig:prore}, the threshold and attention allocation in this figure are reward-rate-optimal. As in that figure, the dashed line marks the `complete forgetting' boundary -- the magnitude of decay beyond which the context statistic decays to zero by the time retrieval starts.}
\label{fig:optattention}
\end{figure}

Figure~\ref{fig:optattention} shows the proactiveness index of the optimal policy of the AX-CPT model for different constraints on memory decay and total drift. We find that, at complete forgetting (high $\lambda$, above the dashed line), the optimal model predicts reactive behavior independent of information processing capacity, consistent with intuition and qualitative framework predictions \cite{Braver2012}. At lower decays, the model behaves proactively in proportion to its drift capacity, behaving more proactively at low total drift and more reactively at high total drift. This provides an argument for proactive behavior as a rational adaptation to the constraints of decaying memory and limited information processing capacity. 

\subsubsection{Discussion}
In this section, we extended our framework to a second task, the AX-CPT, that is different from the Flanker task used in the previous section in a number of important ways: It relies on memory maintenance or retrieval rather than perceptual cue combination, and the response mapping is conditional on, rather than independent of the context. While the Flanker model can be thought of as a continuous-time variant of a previous proposal \cite{Yu2009}, the closest models to the AX-CPT in previous work are the distinct proactive and reactive models of \cite{Lositsky2015}; these be thought of as limiting cases of our AX-CPT model, in which either context or target drift grows large and the model approaches a mixture of two DDMs. %
As with the Flanker model, the AX-CPT model supports a variety of new analyses of processing in this task; for example, the derivation of a Bayes-Risk optimal response and sampling policy, and a reward-rate optimal attention allocation and fixed threshold policy.  The latter provides normative support for the dual mechanisms of control theory \cite{Braver2012}. 

\section{Unification: Simulation of multiple tasks using the same model and parameters}

The previous sections applied our framework to independently to the Flanker and AX-CPT tasks, allowing all parameters to be fit to each individually. Here we consider the extent to which the framework can provide a coherent, unified account of performance \emph{across} tasks, by constraining parameters to be the same for both tasks, except those that define the differences between the two tasks (as detailed below). 
Specifically, we use parameter values previously determined from fits to the Flanker task (from \cite{Yu2009}) apply them without further adjustment to the AX-CPT, except for parameters that reflect structural differences between the tasks.

\subsection{Shared data fitting across tasks}
To model different tasks, we parameterize the response mappings and stimulus onset and offset times according to each task.  Specifically, we assume the model has a high congruence prior for the Flanker model and the correct prior for the AX-CPT (as detailed in Table~\ref{tab:priors}, and uses the memory decay parameter for the AX-CPT but not the Flanker task (which is not applicable in such non-memory tasks). The remaining parameters were constrained to be identical for both tasks, which were taken from \cite{Yu2009} that were used for their model of the Flanker task. For consistency with \cite{Yu2009}, we also added a `premature response' parameter $\gamma$ corresponding to the probability of randomly guessing at time 0. 
We simulated 100,000 trials for each model. 

\begin{figure}[tbh!]
\centering
\includegraphics{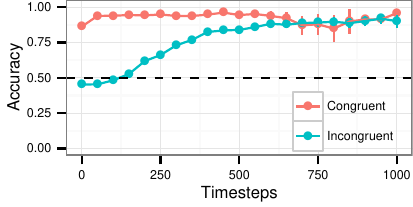}
\includegraphics{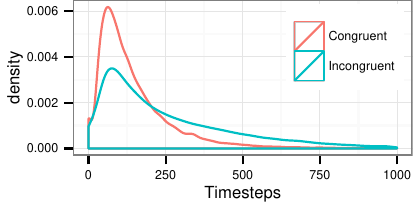}
\caption{\textbf{Unified model recovers characteristic patterns of performance in the Flanker task.} \emph{Left}: response time computed by 50-timestep RT bins for congruent (pink) and incongruent (teal) trials, showing early below-chance performance in the incongruent condition. \emph{Right}: response time distributions for congruent and incongruent trials, showing the same mode but a fatter tail for incongruent relative to congruent trials. Both are signature patterns of performance in the Flanker task, previously reproduced by the Bayesian optimal model of Yu and colleagues.}
\label{fig:Flanker}
\end{figure}

Figure~\ref{fig:Flanker} shows that, for the Flanker task, we recover the characteristic early below-chance performance in incongruent trials. This is an expected verification that our model is a continuous-time limit of the Flanker model of \cite{Yu2009}. 

\begin{table*}[tbh!]
\centering
\begin{tabular}{llllll}
\toprule
   \multicolumn{2}{c}{Context} & \multicolumn{2}{c}{Target} & \multicolumn{2}{c}{Prior} \\
   \cmidrule(l){1-2}\cmidrule(l){3-4}\cmidrule(l){5-6}
   Flanker & AX-CPT & Flanker &AX-CPT  & Flanker &AX-CPT \\
   \midrule
   >\_> & A & > & X & 0.45 & 0.5  \\
   >\_> & A & < & Y & 0.05 & 0.2 \\
   <\_< & B & > & X & 0.05 & 0.2 \\
   <\_< & B & < & Y & 0.45 & 0.1 \\
\bottomrule
\end{tabular}
\caption{Priors for the inference process for the Flanker and AX-CPT instantiation of our theory. }
\label{tab:priors}
\end{table*}

For the AX-CPT task, we compare qualitative patterns of performance in our model with a heterogeneous dataset of humans performing this task ($n=59$) across four different manipulations 
\cite{Lositsky2015} designed to induce ``proactive'' behavior, following \cite{Braver2012}. 

\begin{figure}[tbh!]
\centering
\includegraphics{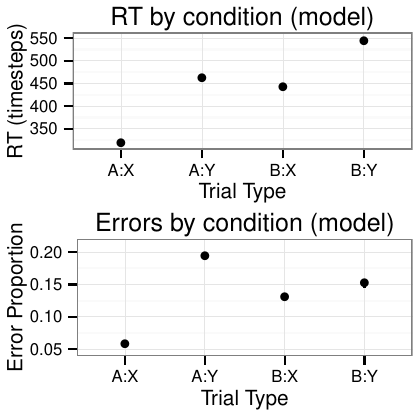}
\includegraphics{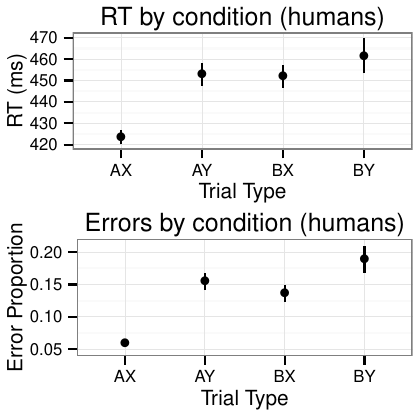}
\caption{\textbf{Unified model recovers qualitative patterns of human performance in the AX-CPT across trial types.} \emph{Left}: RT and error rates by trial type in the model, using the same parameters as the Flanker model. \emph{Right}: RT and error rates by trial type in 59 human participants.
Error bars are standard errors (where not visible, they are smaller than the dots). RT and error patterns observed for the model are qualitatively similar to those observed in humans.
}
\label{fig:axcpt-means}
\end{figure}

\begin{figure}[tbh!]
\centering
\includegraphics[width=0.7\columnwidth]{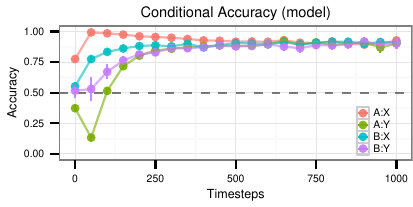}
\includegraphics[width=0.7\columnwidth]{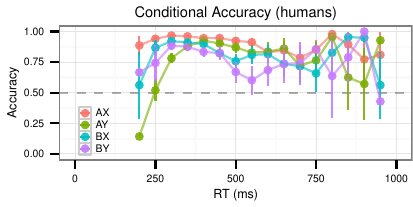}
\caption{ \textbf{Unified model recovers empirically observed pattern in conditional accuracy curves in AX-CPT.} \emph{Left}: response time computed by 50-timestep bin for the four trial types, using the same parameters as the Flanker model. \emph{Right}: same plot from 59 human participants (see text for details). Bins with fewer than 50 observations are omitted. Error bars are standard errors (where not visible, they are smaller than the dots). Both plots show qualitatively similar patterns. Two discrepancies are of note: first, the model predicts very early AY responses to be more accurate than slightly later responses, and early B responses to be close to chance. This may be due to the premature response parameter $\gamma$, which introduces a probability of responding randomly at time 0, disproportionately affecting the accuracy of the earliest responses. We retained it for consistency with the Flanker model. Second, humans show slightly better BY than BX performance early on, something the model does not recover. This may reflect a global left-response bias that the model in the empirical data not captured by the model. %
} 
\label{fig:axcpt-cond-rt}
\end{figure}

Figure~\ref{fig:axcpt-means} shows mean RTs and accuracies produced by the AX-CPT model using the same parameters as the Flanker model. These are qualitatively similar to the pattern observed for human participants. Moreover, the (Figure~\ref{fig:axcpt-cond-rt}) shows that the
model exhibits a pattern in the conditional accuracy curves similar to the one observed for the Flanker task:  below-chance performance for short RTs in AY trials but not other trials. This effect, homologous to the pattern observed at short RTs for incongruent trials in the Flanker task, occurs for the same reason: both are caused by a strong prior biased away from the correct response -- on incongruent trials in the Flanker task, given a high congruence prior; and on AY trials in the AX-CPT, given a high AX prior. 

The results reported above are quantitatively robust to small changes in the prior because eqns.~\ref{eqn:Flanker}~and~\ref{eqn:axcpt-sym} are smooth functions of the prior. The early incongruent errors in Flanker are also robust to larger changes, as long as the congruence prior is above 0.5. The ordering of RTs and error rates for AX-CPT rely on the assumption that participants learn (at minimum) the correct ordering of trial frequencies; this assumption is %
empirically supported \cite{LositskyInPrep}. 

\section{General Discussion}

\subsection{Summary}

In this work, we developed the Contextual Diffusion Decision Model (CDDM), a normative and mechanistic framework for decision making in environments involving multiple sources of information. The CDDM generalizes the classical Diffusion Decision Model (DDM), which provides a normative account of simple two-alternative forced-choice (2AFC) decisions based on a single stream of sensory evidence, to settings in which decisions depend on dynamically evolving contextual and target information. The model is grounded in Bayesian inference, derived as the continuum limit of optimal sequential probabilistic inference over multiple stimuli, and admits a natural implementation as a reduced neural network. In this way, the CDDM provides a unified account linking statistical optimality, behavioral dynamics, and neural mechanisms.

We showed how this framework captures both externally available context, as in the Flanker task, and internally maintained context, as in the AX-CPT and related paradigms. In the Flanker task, we demonstrated that context-dependent biases in decision making emerge naturally from prior statistical dependence between context and target, without requiring additional bias mechanisms. Furthermore, by computing Bayes risk optimal policies, we showed that several previously proposed explanations for Flanker effects, including perceptual uncertainty, and attentional spotlight mechanisms, can be understood as optimal consequences of different assumptions about sampling constraints and costs. Thus, the CDDM provides a normative unification of these seemingly distinct accounts.

In the AX-CPT, the model provides a mechanistic account of context encoding, memory retention, and retrieval as components of a unified inference process. We showed that parameters governing these processes have distinct and interpretable effects on response time distributions, allowing the model to disentangle contributions of encoding, maintenance, and retrieval from behavioral data. Normative analysis further revealed a functional role for memory decay: rather than reflecting a purely detrimental process, decay can mitigate encoding noise and improve decision performance under uncertainty.

Finally, we demonstrated that the CDDM provides a coherent account of behavior across tasks. Using a single set of parameters derived independently from one task, the model reproduced qualitatively accurate patterns of response times and accuracy in another, supporting the claim that the same underlying computational and neural mechanisms may govern diverse forms of context-dependent decision making. These results illustrate the generality of the CDDM as a unified framework for understanding cognitive control and decision making across domains.

Taken together, this work provides a principled extension of diffusion models to multi-stimulus environments, yielding new insights into the normative structure of context-dependent decisions, the mechanistic basis of cognitive control, and the relationship between memory and perceptual inference. The framework opens avenues for future work linking optimal decision theory, neural implementation, and behavioral data in increasingly complex and realistic settings.

\subsection{Future work}

\paragraph{Other tasks} While we have derived specific models for only two tasks, the formalism is applicable to a variety of others. For example, our model of the Flanker could readily be adapted to the Stroop task \cite{Stroop1935} in which, similarly, an irrelevant (and potentially interfering) context is presented concurrently with the target stimulus. The framework could also be applied to paradigms such as the spatial cueing task \cite{Posner1980} in which, like the Flanker (and Stroop) tasks, the context may be congruent or incongruent with the target, but in which the context precedes the target and thus involves a memory component. 

The asymmetric AX-CPT \cite{Henderson2012}, as well as the Prospective Memory task \cite{Einstein2005} would have dynamics similar to the symmetric AX-CPT, but both would share a different response rule in which only one stimulus needs to be distinguished relative to the others. 

Our formalism can also suggest new tasks not yet studied. For example, a congruence judgment task would have the same response rule as the symmetric AX-CPT, but the same stimulus dynamics as the Flanker.

\paragraph{Extensions to the probabilistic formulation} 
As with the conventional DDM, the continuum limit of our model exhibits a scale indeterminacy: multiplying the drift, noise, and threshold by the same constant leaves the predicted response time distributions and choice probabilities unchanged. This occurs because only their ratios determine the evolution of the decision variable~\cite{VS-PH-PS:14e}. Consistent with standard practice, we fix the noise variance to a constant value to make the model identifiable. However, the probabilistic formulation of our model suggests a more principled way to interpret these parameters. In particular, because drift and noise arise from the statistics of stimulus-dependent evidence, they can, in principle, be related directly to properties of the stimulus and sensory encoding, rather than treated as free parameters. This perspective may allow identification of a parameterization in which drift and noise reflect measurable characteristics of the evidence, improving interpretability and facilitating comparison across tasks and conditions (see \cite{Bitzer2014,Brunton2013,Park2016} for related efforts in this direction).

\paragraph{Memory model improvements} Another direction for future work would be to improve the simplistic implementation of context memory in the AX-CPT model. While exponential decay is mathematically convenient, biophysically plausible, rooted in the Ehrenfest urn model, and sufficient to capture essential qualitative features of memory (encoding, retention, retrieval), it is likely that there are additional factors and/or subtleties that remain to be captured by a more elaborate, and potentially realistic model. 
Extensions may take as inspiration belief decay phenomena driven by changing stimuli \cite{Veliz-Cuba2015,Glaze2015}, normative accounts in which memory retrieval dynamically contributes to ongoing evidence accumulation \cite{bornstein2023associative} and attempts to show how a memory system similar to the one we used may arise from simple principles such as uncertainty over stimulus appearance and disappearance times (see also \cite{Meyer2014,Callaway2023}).

\paragraph{Parameter estimation and further cross-task predictions} While we demonstrated qualitatively plausible patterns of performance across two tasks (Flanker and AX-CPT) with a single parameterization, a stronger test would be to generate direct quantitative predictions of behavior in one task based on model parameters estimated from another task, on an individual participant level. Cross-task predictions of this kind would help determine whether the distinct cognitive tasks that we consider in a unified framework indeed tap the underlying computationally precise cognitive constructs implemented by our models. Doing so would require a robust method for parameter estimation. In contrast to the DDM, which is a linear diffusion model to fixed boundaries, our model's likelihood is the first passage time of a 1D nonlinear diffusion model to a fixed boundary, or equivalently, the first passage time of a 2D linear diffusion model to a nonlinear boundary. In either formulation, standard expressions for the first passage times of diffusion models are not applicable, and the first passage time is determinable only by solving a partial differential equation describing the flow of the probability mass of the context and target statistics. While such an approach has been tried for the DDM \cite{Voss2008}, the boundary shapes given in Figure~\ref{fig:cddm-2d} imply an infinite solution region, which requires nontrivial PDE solution methods that we found to be difficult to embed in a stable way in the inner loop of a parameter estimation routine. A more promising approach may be to treat the model as a fully likelihood-free simulation model, and apply one of a variety of likelihood-free inference methods for parameter estimation \cite[e.g.][]{Barthelme2011,fengler2021,tejero-cantero2020}. 

\paragraph{Neural evidence} The DDM has received neuroscientific support from evidence of rise-to-threshold neural activity in a variety of population recordings \cite[e.g.][]{Kira2015,Gold2002,Gold2007Neuro,Resulaj2009,Hanks2015}. In contrast to the linear average trajectories of the DDM, the response statistics (Figures~\ref{fig:temporally-seperated-stimuli}~and~\ref{fig:Flanker-decomposed}) and stopping policies (Figures~\ref{fig:flanker-boundaries}~and~\ref{fig:cddm-2d}) of models in our framework have distinct nonlinear shapes that could be compared to, and may capture more nuanced features of neural data. In addition to testing the dynamic predictions of our model, localization of neural components may address questions we presently do not address, such as whether the dynamic weighting happens at the sampler or further upstream (i.e.\ whether unreliable evidence is gated at the sampler or discounted at the integrator).

\subsection{Conclusions}
We have provided a theoretical framework for understanding decision making involving the contextual influence on decision making, both when the context is relevant or irrelevant (but nonetheless potentially interfering), and when the context precedes or is presented concurrently with the target. We have shown how the formalism can be derived both from a biophysically-inspired connectionist model and a probabilistic inference perspective. Our work generalizes similar joint motivations of the DDM, the dominant model of two-alternative decision making in response to a single stimulus. We used our theory to derive models of two distinct tasks from the cognitive psychology literature involving context and target stimuli, one a notational equivalent of a previous model (of the Flanker task) and the other a novel model of a simple, well-established context-dependent decision making task (the AX-CPT). We showed how we can write these models in terms of nonlinear transformations of constant-drift random walks, and computed optimal policies for both, in both cases with detailed psychological interpretations. Finally, we showed how models derived from our framework can be used to generate sensible behavioral predictions without parameter changes across tasks. These findings extend previous efforts to identify fundamental, formal, rigorous principles of decision making to address both the dynamics and the influence of context processing — a feature that is central to most real-world behavior.  We hope that this work will help lay the foundation for more complex, realistic models of such behavior.

\pagebreak

\bibliographystyle{ieeetr}
\bibliography{mike,cddm}

\onecolumn
\appendix

\pagebreak

\pagebreak

\section{O-U process as a continuum limit of a birth-death Markov chain} \label{sec:ehrenfest}

Here we provide a new mechanistic motivation for the O-U process as a model of memory. Specifically, we assume that the memory trace is maintained as a set of veridical stimulus samples corrupted by a set of noise samples and that at every time step a veridical sample may be replaced with a noise sample. We map this process to the Ehrenfest Urn model in which there is an urn containing $n$ balls, and the balls can be of two colors (for us, the correct memory or or a noise sample). 
At each time, one of the balls (samples) is selected at random. If the selected sample is the true memory, then it is replaced by a noise sample with probability $\alpha>0$. Likewise, a noise sample can be replaced with a veridical sample with probability $\beta>0$. Let the number of veridical memory samples in the urn at time $k$ be a random variable $Y_k$. Then, conditioned on $Y_k=x$, the probability distribution of $Y_{k+1}$ is 
\[
P(Y_{k+1} = y-1 | Y_k=y) = \frac{\alpha y}{n}, \quad   P(Y_{k+1} = y+1 | Y_k=y) = \frac{\beta (n-y)}{n},
\]
and $P(Y_{k+1} = y | Y_k=y) = 1 - P(Y_{k+1} = y-1 | Y_k=y) - P(Y_{k+1} = y+1 | Y_k=y)$.

The above birth-death chain admits a unique stationary distribution \cite{durrett1999essentials}, which is a binomial distribution with a trial size $n$ and success probability $\beta /(\alpha + \beta )$. The associated mean and variance are $ n\beta/(\alpha + \beta )$ and $n \alpha \beta  /(\alpha + \beta )^2$, respectively.

We now focus on constructing the continuum limit of the above Ehrenfest urn model. It can be seen in the above expressions that, unlike a random walk, the state transition matrix of the Ehrenfest urn model is state-dependent.  Therefore, the standard arguments~\cite{Bogacz2006} that construct the DDM as a continuum limit of a random walk can not be used. Instead, we rely on the theory of convergence of Markov chains to diffusion process~\cite[Section 8.7]{durrett1996stochastic}. 

To construct the continuum limit, let $\Delta t = \frac{1}{n}$, and consider the continuous time random variable 
\[
X_t^{(n)} = \frac{Y_{\lfloor{nt}\rfloor} - \frac{n \beta}{(\alpha +\beta)}}{\sqrt{n}}, 
\]
where $\lfloor\cdot\rfloor$ represents the largest integer smaller than or equal to the argument, $X_t^{(n)}$ corresponds to the state of the above birth-death chain after $\lfloor nt\rfloor$ time steps centered using steady-state mean and scaled using steady-state standard deviation. 

Note that $X_t^{(n)}=x \implies Y_{\lfloor{nt}\rfloor} = x \sqrt{n} + \frac{n \beta}{(\alpha +\beta)}$. Therefore, it follows
\begin{align*}
    P\left(X_{t+\Delta t}^{(n)} = x- \frac{1}{\sqrt{n}} \Big| X_{t}^{(n))}=x\right) &= \alpha  \frac{x \sqrt{n} + \frac{n \beta}{(\alpha +\beta)}}{n} \\
        P\left(X_{t+\Delta t}^{(n)} = x- \frac{1}{\sqrt{n}} \Big| X_{t}^{(n))}=x\right) &= \beta  \frac{-x \sqrt{n} + \frac{n \alpha}{(\alpha +\beta)}}{n}.
\end{align*}

Following Theorem 7.1 and Example 8.1 in~\cite{durrett1996stochastic}, we define 
\begin{align*}
    b^{\Delta t}(x) &= \frac{1}{\Delta t}\mathbb{E}[X_{t + \Delta t}^{(n)} - X_t^{(n)}| X_t^{(n)}=x] = -(\alpha + \beta) x \\
     a^{\Delta t}(x) &= \frac{1}{\Delta t}\mathbb{E}[(X_{t + \Delta t}^{(n)} - X_t^{(n)})^2| X_t^{(n)}=x] = \frac{(\alpha -\beta)}{\sqrt{n}} x + \frac{2 \alpha \beta}{\alpha + \beta}
\end{align*}
Taking the limit $n \to \infty$ (equivalently, $\Delta t \to 0$), we have
\[
b^{\Delta t}(x) \to b(x) = -(\alpha + \beta) x , \quad \text{and} \quad  a^{\Delta t}(x) \to a(x) = \frac{2 \alpha \beta}{\alpha + \beta}. 
\]
Using Theorem~7.1 in~\cite{durrett1996stochastic}, in the limit $n \to \infty$, $X_t^{(n)}$ converges to the following O-U process
\[
d X_t = - (\alpha + \beta ) X_t dt + \sigma d W_t,
\]
where $\sigma^2 = \frac{2 \alpha \beta}{\alpha + \beta}$. Note that the steady-state distribution of this O-U process is Gaussian with zero mean and variance $\frac{\sigma^2}{2(\alpha+\beta)} = \frac{\alpha \beta}{(\alpha +\beta)^2}$. This variance is consistent with the steady-state variance of the Ehrenfest urn model discussed earlier, which interprets $X_t$ as the fraction of veridical memory samples in the urn. This scaling by $n$ is due to the division by $\Delta t$ in the definitions of $b^{\Delta t}(x)$ and $a^{\Delta t}(x)$. The above discussion considered the zero-mean O-U process. We can include the steady-state mean of the Ehrenfest urn model to determine the non-centered O-U model
\begin{align*}%
    d X_t = (\alpha + \beta ) \left(\frac{\beta}{\alpha + \beta}  -  X_t \right)dt + \sqrt{\frac{2 \alpha \beta}{\alpha+\beta}} d W_t. 
\end{align*}
When $\beta$ is small, i.e., the green balls are not replaced with red balls, the constant term in the drift is negligible, and the noise is also insignificant. Thus, the model acts as an exponential decay model. This regime is consistent with waiting for the target stimulus phase of the task, where no stimulus is present, and the decision maker slowly forgets the context. When $\beta$ is large, the constant term of the drift takes a sufficiently large value. This regime is consistent with the decision making phase of the task, where the decision maker recalls the context from memory and replaces noise samples with veridical samples.

\end{document}